\RequirePackage{fix-cm}
\documentclass[twocolumn,epjc3]{svjour3}  
\smartqed  
\RequirePackage{graphicx}
\journalname{Eur. Phys. J. C}

\usepackage{xcolor}
\usepackage{color}
\usepackage{epsf}
\usepackage{soul}
\usepackage{etex}
\usepackage{morefloats}

\colorlet{darkgreen}{green!50!black}
\colorlet{brightyellow}{yellow!75!red}
\colorlet{orange}{red!50!yellow}
\colorlet{darkblue}{blue!60!black}
\colorlet{darkred}{red!80!black}

\newcommand{\cd}{\makebox[0.08cm]{$\cdot$}}

\begin{document}

\title{Hybrid nature of  the abnormal solutions of the Bethe-Salpeter
equation in the Wick-Cutkosky model}

\author{J.~Carbonell\thanksref{e1,addr1}
\and
V.A.~Karmanov\thanksref{e2,addr2}
\and
H.~Sazdjian\thanksref{e3,addr1}
}
\thankstext{e1}{e-mail: carbonell@ipno.in2p3.fr}
\thankstext{e2}{e-mail: karmanovva@lebedev.ru}
\thankstext{e3}{e-mail: sazdjian@ipno.in2p3.fr}

\institute{Universit\'e Paris-Saclay, CNRS/IN2P3, IJCLab,
91405 Orsay, France\label{addr1}
\and
Lebedev Physical Institute, Leninsky prospect
53, 119991 Moscow, Russia\label{addr2}
}

\date{Received: date / Accepted: date}

\maketitle

\begin{abstract}
In the Wick-Cutkosky model, where two sca\-lar massive constituents
interact by means of the exchange of a scalar massless particle,
the Bethe-Salpeter equation has solutions of two types, called
``normal'' and ``abnormal''.
In the non-relativistic limit, the normal solutions correspond to the
usual  Coulomb spectrum, whereas the abnormal ones do not have
non-relativistic counterparts -- they are absent in the
Schr\"o\-din\-ger equation framework. 
We have studied, in the formalism of the light-front dynamics,  the
Fock-space content of the abnormal solutions. 
It turns out that, in contrast to the normal ones, the abnormal states
are dominated by the massless exchange particles (by 90 \% or more),
what provides a natural explanation  of their decoupling from the
two-body Schr\"odinger equation.
Assuming that one of the massive constituents is charged, we have
calculated the electromagnetic elastic form factors of the normal
and abnormal states, as well as the  transition form factors.
The results on form factors confirm the many-body nature of the abnormal states, as
found from the Fock-space analysis.
The abnormal solutions have thus properties similar to those of
hybrid states, made here essentially of two massive constituents and
several or many massless exchange particles.
They could also be interpreted as the Abelian scalar analogs of the
QCD hybrid states. The question of the validity of the ladder
approximation of the model is also examined.
\end{abstract}

\section{Introduction}\label{Intro}

In their pioneering papers \cite{wick,cutk}, Wick and Cutkosky (W-C) have
found the solutions of the Bethe-Salpeter (BS) equation \cite{bs} for
two scalar particles interacting by the exchange of a massless scalar
particle. In addition to the states which, in the non-relativistic
limit, reproduce the spectrum of the Schr\"odinger equation with the
Coulomb potential, there was found another set of solutions which do
not have any non-relativistic counterparts. These solutions were
called ``abnormal''.
Their discovery triggered the discussion as to whether they do
indicate a mathematical inconsistency of the W-C model or
of the BS equation, or whether they represent new physical systems,
whose existence does not contradict any physical principles, 
although they are not covered by the Schr\"odinger equation.
In the latter case, they might provide examples of relativistic
systems which could exist in nature, but which would not be
described by continuous extensions of non-relati\-vis\-tic quantum
mechanics. A thorough discussion of this issue can be found in Ref.
\cite{nak69} (sections 6 and 8).

One would hope that a complementary lighting to the above questioning
might come from experimental data. Unfortunately the conditions of the
emergence of abnormal states are not easy to realize.
Considering the W-C model as a simplified model of QED,
abnormal states would appear as highly excited states for values  of
the fine
structure constant $\alpha$ above $0.5\div 1$. Such values might be
reached with the
aid of heavy ions and dedicated electron-ion scattering experiments
might be envisaged. However, to have a clear experimental distinction
of abnormal bound states from the ionization threshold, one actually
would need to increase the values of $\alpha$ up to $4\div 5$, which
then further reduce the probability of an experimental success.
Another possibility is an analogy of the model with hadron dynamics,
where hadrons mutually interact by means of the exchange of light
particles, like the pions, and where the coupling constants might
lie in the range of values needed for the existence of abnormal
states. However, here, the exchanged particles being massive,
drastic changes occur with respect to the massless case: the
interaction forces become of short-range and one realizes that
abnormal states are produced only with very small mass values of
the exchanged particle, much smaller than the pion mass. 
The framework of Quantum Chromodynamics, where quarks
mutually interact by means of exchange of massless gauge particles,
the gluons, with sufficiently strong forces, might provide another
domain to search for  possible evidences of abnormal solutions.

In the absence of any direct experimental indication about the
existence or nonexistence of abnormal states, one is entitled to
explore all possible theoretical paths that might provide
complementary information about their properties. From this point
of view, we have found that an analysis of the Fock-space content
of the abnormal, as well as normal, states would be of great help.
The BS amplitude allows one to extract the wave function related
to the two-body sector of the Fock space \cite{cdkm}. 
Its norm, which is positive and bounded by 1, is then
interpreted as the weight of that sector in the whole Fock space.

Complementary information to the above analysis comes from the
knowledge of the electromagnetic form factors, assuming that one
of the massive constituents of the bound state is charged. Their
asymptotic behavior qualitatively probes the compositeness of the
states: a rapid decrease would be the signature of a many-body
structure \cite{matvmurtavk,brodsfarr,radyush}. 

Our calculations, as well as the results of ref. \protect{\cite{dshvk}},
show that, in the window of allowed  coupling constants,  the normal
solutions are essentially dominated by the two-body sector of the Fock
space. 
We will show in the present work that, on the contrary, 
the abnormal solutions have a two-body contribution that vanishes in
the limit of zero binding energies and remains small (less than
10\%) in all  its domain of existence.
They are therefore dominated by the many-body sectors, composed of the
two massive constituents and of several massless exchange
particles. This feature explains why the abnormal
solutions disappear from the spectrum in the non-relativistic
limit, the latter being formulated in the two-body sector alone,
while the other sectors, containing massless particles, are by
essence relativistic.

The asymptotic behaviors of the form factors also corroborate the
above conclusions. The form factors of the abnormal solutions
asymptotically decrease, for spacelike momenta, faster, by factors
of the order of $10^{3}$, than those of the normal solutions.
Also, the transition form factors between normal and abnormal
solutions display global suppressions, by factors of $5\div 10$, with
respect to the normal-normal or abnormal-abnormal transition form
factors, signalling a different nature of  the normal and abnormal
solutions.

These results suggest that the abnormal solutions might correspond
to states called ``hybrids'' in the literature. In the present model,
they are dominated
by  Fock space sectors containing  two massive constituents and several
or many massless constituents, corresponding to the exchanged-particle
fields.
They could be considered as the Abelian scalar analogs of the QCD
hybrids, which, in the mesonic sector, are dominated by their
coupling to the set of fields made of a quark, an antiquark and one or
several gluon fields.

Finally, the question of the validity of the ladder approximation
of the model, because of the necessity of having large values of the
coupling constant to create abnormal states, still remains an open issue.

The plan of the paper is the following. Sec.  \ref{def} is devoted to an
introductory definition of the BS amplitude and of the Fock-space sectors. 
In Sec. \ref{WCsol}, the properties of the solutions of the
W-C model are displayed and some solutions are found numerically. 
In Sec. \ref{FFs} the elastic and transition electromagnetic form factors
are expressed through the BS amplitudes and are calculated numerically,
with special emphasis put on their asymptotic behavior. Concluding remarks
follow in Sec. \ref{concl}. 
Three appendices give technical details
about some of the formulas used in the main text. Preliminary results
of the present study were presented in Ref. \cite{LC2019}.

\par
\section{Fock space sectors} \label{def}

The BS amplitude, satisfying the BS equation, is defined as
\begin{equation}\label{bs1}
\Phi(x_1,x_2;p)\equiv \langle 0 |T[\phi_1(x_1)\phi_2(x_2)]|p\rangle,
\end{equation}
where $\phi_a(x)$ ($a=1,2)$ are Heisenberg field operators,
$T$ means time ordering, $|p\rangle$ is the state vector of the
bound system and $\langle 0 |$ is the vacuum state vector.
Since the amplitude  $\Phi(x_1,x_2;p)$ depends on two 4D variables,
$x_1$ and $x_2$, it is usually called ``two-body'' BS amplitude,
though this terminology, to some extent, is misleading.
Asking the questions ``what is the content of a system?'' or ``is it
two-body or many-body?'' requires that we analyze the state
vector $|p\rangle$ of this system, entering in the  matrix element
(\ref{bs1}), by decomposing it onto the states $| n\rangle$ with
definite numbers $n$ of particles (the Fock sector decomposition),
schematically:
\begin{equation}\label{p}
|p\rangle=\sum_{n=2}^{\infty}\psi_n^{}|n\rangle,
\end{equation}
and studying the contributions of the two-body component $\psi_2$,
the three-body component $\psi_3$, etc., in the full normalization
integral. The answer to the above questions depends on which component
(or sum of components) is dominant.  It should be mentioned that the state vector $|p\rangle$
  is usually defined on a  $t$-constant plane in the 4D space. There are however some advantadges to choose the so called
light-front plane 
$t+z=0$ (or light-front plane of general orientation, see \cite{cdkm}). In this case, the corresponding Fock components 
$\psi_n$ are called the light-front wave functions. The two-body light-front wave function $\psi_2$ is related to the BS 
amplitude (\ref{bs1}) by eq. (\ref{lfwf}) from \ref{appN2}.
\par

In the W-C model, in the ladder approximation, the
Fock decomposition can contain two constituent (massive) particles
and any number of exchange (massless) particles. The state
$|2\rangle$ (two-body sector) contains two constituents only, the
state $|3\rangle$  (three-body sector) contains two constituents and
one exchange particle, the state $|n\rangle$  ($n$-body sector)
contains two constituents and $(n-2)$ exchange particles, etc. 
Assuming that the state vectors $|n\rangle$ ($n=2,3,\ldots$) are
normalized to unity, the state vector $|p\rangle$ is then normalized
as
\begin{equation}\label{eq2}
\langle p|p\rangle=\sum_{n=2}^{\infty}N_n=1,
\end{equation}
where, schematically, $N_n=\int |\psi_n|^2\ldots$ 
is the contribution of the $n$-body Fock sector
(see eq. (\ref{N2}) for the exact definition of $N_2$).
In practice, knowing the BS amplitude $\Phi(x_1,x_2;p)$ we are able to find $N_2$ only.  
The calculation of $N_2$ from the BS amplitude is presented in
\ref{appN2}. If it is dominant, this would mean that the
contribution of the other sectors, containing exchange
particles, is small. The limiting case, when it is enough to keep the
two-body state only (the case corresponding to $N_2=1$), whereas the
states containing exchange particles can all be omitted, is realized
in non-relativistic systems. On the contrary, when the two-body
contribution $N_2$ is small, the system is dominated by 
two constituents with an indefinite number of exchange massless 
particles, whose contribution $\sum_{n=3}^{\infty}N_n$ is close to 1.
\par

For the normal solutions of the W-C model (in the
equal-mass case), the above analysis
has been made in Ref. \cite{dshvk}. It was found that for small binding
energies the two-body (constituent) sector dominates, as expected. When
the binding energy increases (i.e., the total mass $M$ decreases), the
two-body contribution $N_2$ decreases in parallel. However, it still
dominates and as $M\to 0$, it tends, in this model, to 64\%.
That is, the sectors  $|n\rangle$ with $n\geq 3$  contribute in total
to 36\% of the total normalization of the normal state vector.
In the present paper, we will carry out the same analysis for the
abnormal states.

\section{Wick-Cutkosky solutions}\label{WCsol}

The BS equation \cite{bs} for the amplitude (\protect{\ref{bs1}})
containing two spinless fields, 
restricted to the equal-mass case $m_1=m_2=m$, reads, in momentum space,
\begin{eqnarray}\label{bs}
\Phi(k,p)&=&\frac{i^2}{\left[(\frac{p}{2}+k)^2-m^2
+i\epsilon\right]\left[(\frac{p}{2}-k)^2-m^2+i\epsilon\right]}
\nonumber\\
&\times&\int \frac{d^4k'}{(2\pi)^4}iK(k,k',p)\Phi(k',p),
\end{eqnarray}
where $p$ and $k$ are the total and relative four-momenta,
respectively, and $K$ is the interaction kernel. The bound state
mass squared is $M^2=p^2$. For nonconfining interactions, the mass $M$
is smaller than $2m$, allowing the introduction of the binding energy
$B$ (defined positive) through the relation $M^2=(2m-B)^2$.
In the ladder approximation of the kernel, represented by the exchange
of a scalar particle with mass $\mu$, the kernel has the form
\begin{equation}\label{ladder}
iK(k,k',p)=\frac{i(-ig)^2}{(k-k')^2-\mu^2+i\epsilon},
\end{equation} 
leading to an attractive interaction and the possible emergence of
bound states.

\subsection{General properties of the solutions} \label{genprop}

The W-C model corresponds to the case
$\mu=0$ in Eq. (\ref{ladder}). In the non-relativistic limit, this
model leads to the well-known Coulomb bound state spectrum.
Cutkosky showed that 
in the relativistic case,
the BS amplitude,
henceforth limited to $S$-wave states, characterized by a principal
quantum number $n=1,2,\ldots$, can be represented  in terms  of $n$
functions  $\left\{ g_n^{\nu} \right\}_{{\nu}=0,1\ldots n-1}$, depending on
a single scalar argument  $z\in[-1,+1]$,   as 
\begin{eqnarray}\label{Phi}
&& \Phi_n(k,p)=- {i \over \sqrt{N_{tot}}}
\sum_{{\nu}=0}^{n-1}\int_{-1}^1g_{n}^{\nu}(z)dz  \nonumber \\
&& \times\frac{m^{2(n-{\nu})+1}}
{[m^2-\frac{1}{4}M^2 -k^2-p\cd k\,z-\imath\epsilon]^{2+n-{\nu}}},
\ n=1,2,\ldots\ .    \nonumber \\
&&       
\end{eqnarray}
$N_{tot}$  is a dimensionless normalization factor, deter\-min\-ed in
\ref{exprssff},
ensuring the condition $F_{el}(0)=1$  for the elastic form factor . 
The factor $m^{2(n-{\nu})+1}$ in the numerator is introduced to deal
with dimensionless  $g_{n}^{\nu}(z)$  functions.

By inserting (\ref{Phi}) in the  BS equation (\ref{bs}), Cutkosky
obtained [Eq. (14) of Ref. \cite{cutk}] a system of homogeneous coupled
integral equations for the functions $g_n^{\nu}$.
For S-waves, it reads\footnote{A typo seems to exist in Eq. (14) of
Ref. \cite{cutk}: the integration with respect to $t$ goes from $-1$
to $+1$, and not from $0$ to $+1$, as can be verified from Eq. (13)
of that reference. Notice that we use a slightly different notation, 
replacing $k \to \nu$, with respect to the original work \cite{cutk}.}:
\begin{small}
\begin{eqnarray}\label{ECut}
&&g_n^{\nu}(z)= {\lambda\over2}  \sum_{{\nu}'=0}^{\nu} \frac{ (n-{\nu}+1)(n-{\nu})}
  {  (n-{\nu}'+1) (n-{\nu}')}   \int_{-1}^1dt  \int_0^1 dx\nonumber   \\
  &\times&       x(1-x)^{n-{\nu}-1} \int_{-1}^{+1}dz'
  \frac{\delta[ z -xt-(1-x)z'] }{ [1-\eta^2(1-z^2) ]^{{\nu}-{\nu}'+1}}
  g_n^{{\nu}'}(z'),   \nonumber\\
&&  
\end{eqnarray}
\end{small}
where $\lambda$ is related to the coupling constant $g^2$ of the
interaction kernel  (\ref{ladder}) by
$\lambda=  {g^2\over 16 \pi^2m^2}, $
and the total mass square $M^2$, eigenvalue of the system (\ref{ECut}),
appears through the parameter
\[  \eta^2={M^2\over 4m^2}.  \]
Integrating Eq. (\ref{ECut}), first with respect to $t$ through
the $\delta$ function, taking into account the bounds to be
satisfied by $t$ and $x$, and distinguishing the two cases, $z>z'$
and $z'>z$, one obtains the set of equations
\begin{eqnarray}\label{ECut2}
  g_n^{\nu}(z)&=&  {\lambda\over2}  \sum_{{\nu}'=0}^{{\nu}}   c_n^{{\nu}{\nu}'} \;
  \int_{-1}^{+1} dz' \; {[R(z,z')]^{n-{\nu}} \over [Q(z')]^{{\nu}-{\nu}'+1}} \;
  g_n^{{\nu}'} (z'), \nonumber   \\
  && n=1,2,\ldots , \; \; {\nu}=0,1,...n-1,   
\end{eqnarray}
where we have introduced
\[ c_n^{{\nu}{\nu}'}=\frac{ (n-{\nu}+1)} {  (n-{\nu}'+1) (n-{\nu}')},  \]
\begin{equation}\label{Q}
 Q(z)=1-\eta^2(1-z^2), 
 \end{equation}
and
\begin{equation}\label{eq1bb}
R(z,z')=\left\{
\begin{array}{ll}
\frac{1-z}{1-z'},& \mbox{for $z'<z$,}
\vspace{0.2cm}
\\
\frac{1+z}{1+z'},& \mbox{for $z'>z$.}
\end{array}
\right.
\end{equation}
By expanding Eq. (\ref{ECut2}), one is left with  a $n\times n$
triangular system of one-dimensional integral equations of the form
\begin{small}
\begin{eqnarray}
g_n^0(z) & =& {\lambda\over2}
  \left[ c_n^{00} \int_{-1}^{+1} dz' \;  { [R(z,z')]^n \over Q(z')}
    \; g_n^0  (z') \right],  \label{ECut2_0}  \\
g_n^1(z) &=& {\lambda\over2}
  \left[ c_n^{10} \int_{-1}^{+1} dz' \; {[R(z,z')]^{n-1} \over [Q(z')]^2}
    \; g_n^0  (z')  \right.
  \nonumber   \\
    &+&  \left.c_n^{11} \int_{-1}^{+1} dz' \;
       {  [R(z,z')]^{n-1} \over Q(z') } \; g_n^1  (z')  \right], 
       \nonumber\\
g_n^2(z) & =& {\lambda\over2}
  \left[ c_n^{20} \int_{-1}^{+1} dz' \; {[R(z,z')]^{n-2} \over [Q(z')]^3}
    \; g_n^0  (z')  \right.
    \nonumber\\
     &+& \left. c_n^{21} \int_{-1}^{+1} dz' \;
       {[R(z,z')]^{n-2} \over [Q(z')]^2 } \; g_n^1  (z')\right.
       \nonumber \\
&+&\left. c_n^{22} \int_{-1}^{+1} dz' \;
       {  [R(z,z')]^{n-2} \over Q(z') } \; g_n^2  (z')\right],  
       \nonumber\\   
& &  \ldots  \ \ \ \ \ \ \ \ldots \ \ \  \; \ldots \ \ \  \ldots  \;
  \ \ \ \ldots\ ,   
  \nonumber\\
g_n^{n-1}(z) &=& {\lambda\over2}
  \left[  c_n^{n-1,0}   \int_{-1}^{+1} dz' \; { R(z,z') \over [Q(z')]^n}
    \; g_n^0  (z')   +  \ldots \right. 
    \nonumber\\
    &+&   \left.c_n^{n-1,n-1}
    \; \int_{-1}^{+1} dz' \;    { R(z,z') \over Q(z')}  \; g_n^{n-1}(z')
    \right].  \nonumber
\end{eqnarray}
\end{small}

Remarkably, the function $g_n^0$, which allows the calculation of the
energy spectrum via the $M^2$-dependence  of $Q$ [Eq. (\ref{Q})],  is
totally decoupled from the rest of the system.
It fulfills the single equation (\ref{ECut2_0}), that we will hereafter
write in terms of the fine structure coupling constant $\alpha$, usual
in the Coulomb problems:
\begin{eqnarray} \label{gn}
  g_n^0(z)&=&\frac{\alpha}{2\pi n}  \int_{-1}^1 {[R(z,z')]^n \over Q(z')}
  \; g_n^0  (z'),  
  \\
  \alpha&=&\pi \lambda = {g^2\over 16\pi m^2}.
  \nonumber
\end{eqnarray}

The remaining equations allow the determination of  $g_n^{k>0}$ --  and
so of the BS amplitude (\ref{Phi}) -- by solving an inhomogeneous
problem with  an inhomogeneous term given by $g_n^0$.
Notice that it is a quite unusual situation in Quantum Mechanics that
a part of the total system wave function,  which, as we will see in what
follows is far from being dominant, 
determines the full spectrum of the system.

Although the results presented here are limited to  $S$-wave only, it is worth noticing  that for $l\neq 0$
the corresponding spherical function $Y_{lm}$ would appear as a prefactor
in Eq. (\ref{Phi}). 
The angular momentum  $l$ would enter in the system of equations (\ref{ECut2}) and (\ref{ECut2_0}), 
but it turns out to be absent in the first equation  (\ref{ECut2_0}) determining the spectrum.
As a consequence, the BS amplitude would depend on $l$, while  the spectrum would remain  $l$- degenerate. 

In view of its numerical solution, it is interesting to write Eq.
(\ref{gn})  in a differential form:
\begin{eqnarray}\label{gndf}
&&g_n^{0\,\prime\prime}(z)+2(n-1)z(1-z^2)^{-1}g_n^{0\,\prime}(z)
\nonumber \\
&&\ \ \ \ \ \ \ \ \ -n(n-1)(1-z^2)^{-1}g_n^0(z)\nonumber \\
&&\ \ \ \ \ \ \ \ \ +\frac{\alpha}{\pi}\frac{1}{(1-z^2)Q(z)}g_n^0(z)=0,
\end{eqnarray}
with the boundary conditions $g_n^0(\pm 1)=0$.
\par

For a fixed $n$, Eq. (\ref{gndf})  has an infinite number of
solutions, labeled by  an additional quantum number
$\kappa=0,1,2,\ldots$, which also labels the corresponding discrete
spectrum of mass squared eigenvalues $M_{n\kappa}^{2}$. 
We will use the notation $g^{\nu}_{n\kappa}$ to identify a particular
solution.
The function {$g_{n\kappa}^0$} has $\kappa$ nodes within the interval
\mbox{$]-1,+1[$}  and a well-defined  parity given by $\kappa$ \cite{cutk}:
\[g^0_{n\kappa} (-z)= (-1)^{\kappa} g^0_{n\kappa}(z) \]
The parity  is also preserved inside the  ensemble  $\{ g_{n\kappa}^{\nu} \}$
when varying $\nu=0,1,\ldots$  and this entails, through Eq. (\ref{Phi}),
that for even (odd) values of $\kappa$,
the BS amplitude $\Phi(k,p)$ is an even (odd) function of the relative
energy $k_0^{}$ in the c.m. frame. \par

\begin{figure}[h!]
\begin{center}
\mbox{\epsfxsize=8.cm\epsffile{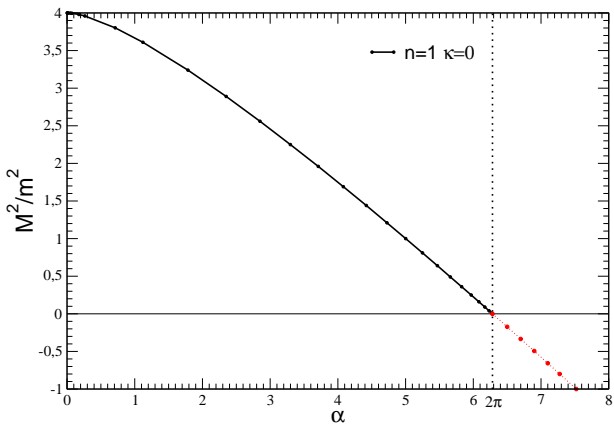}}  
\hspace{.5cm}
\end{center}
\caption{(Color online) Dependence  of the squared mass of the ground
  state ($n=1, \kappa=0)$ on the coupling constant $\alpha$. 
  Beyond the critical value $\alpha=2\pi$, the solutions are smootly
  continued without any singularity but having  negative values of
  $M^2$. 
  The physical region, where the system has well-defined ground state
  mass and binding energy, is thus limited to $\alpha\in[0,2\pi]$.}
\label{Fig_M2_alpha_10}
\end{figure}

The mass squared $M^2$ of the ground state $g_{10}^0$ as function of the
coupling constant  $\alpha$ is shown in Fig. \ref{Fig_M2_alpha_10}.
Its value vanishes for $\alpha=2\pi$. In the range $\alpha \in [0,2\pi]$,
$M^2\ge 0$, the total mass of the system $M$ is well  defined as well as
its binding energy  $B=2m-M>0$.
This determines the domain where this model is physically consistent
with a well-defined ground state.
It is worth mentioning, however, that the solutions of the BS equation,
as well as its spectral parameter $M^2$, can be analytically continued
for $\alpha>2\pi$ 
without encountering any kind of singularity. This is illustrated with
the  dashed line in the lower right corner of the figure.  
All excited states lie above the $M^2(\alpha)$ curve and thus can have
a well-defined $M$ even in the unphysical region.

 \vspace{.5cm}

Among the infinity of solutions existing for a given  $n$, the one with
$\kappa=0$ coincides,  in the  limit 
of small binding energies $B/m\ll 1$, with the solution of the
non-relativistic Coulomb problem with main quantum number $n$.  
This solution is called, following the original works of
Wick and Cutkosky \cite{wick,cutk},  ``normal''.
Indeed, these authors, analyzing in this limit the system of equations
for the functions $g_n^k(z)$ determining the BS amplitude (\ref{Phi}), 
reproduced, for $\kappa=0$, the Coulomb spectrum, i.e. the Balmer series 
\begin{equation}\label{Balmer}
B_n=\frac{m\alpha^2}{4n^2}.
 \end{equation}
This result corresponds to the Schr\"odinger equation with the potential
$V(r)=-\frac{\alpha}{r}$. 
The relativistic perturbative correction to
the binding energy (\ref{Balmer}) was found in \cite{FFT}; the binding
energy, incorporating it, reads
\begin{equation}\label{B_Pert}
  B_n=\frac{m\alpha^2}{4n^2}\left[ 1 -    {4\alpha\over\pi}
    \ln\left({1\over\alpha}\right)  + {\mathcal O}(\alpha) \right]. 
  \end{equation}

On the contrary,  the solutions  corresponding to non-zero  values of
$\kappa$ ($\kappa=1,2,\ldots$),   have a spectrum  totally decoupled
from the non relativistic one.
They are genuinely of relativistic nature, without non-relativistic
counterparts, and were named  "abnormal" by Wick.

\begin{figure}[h!]
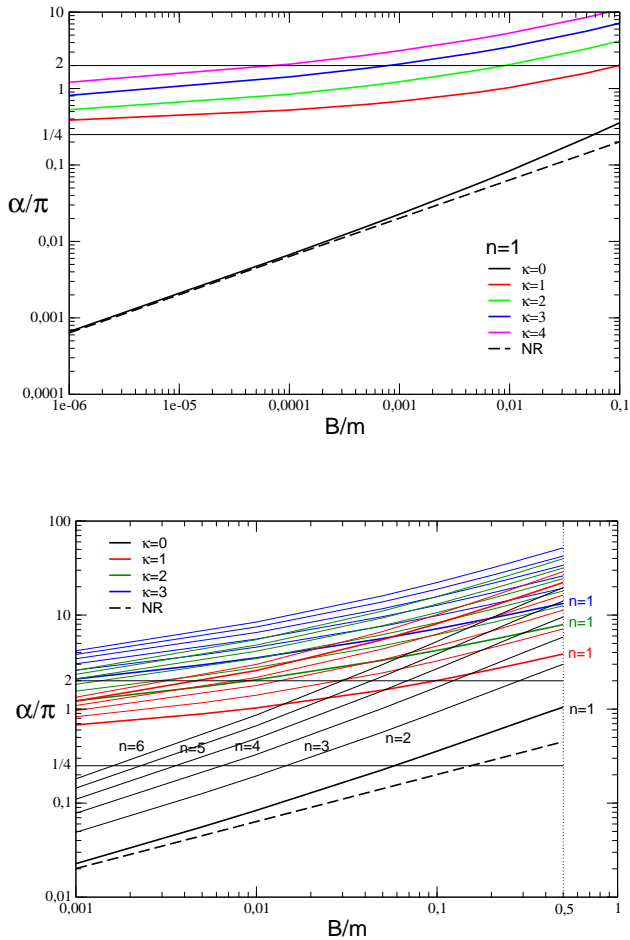

\vspace{.5cm}
\begin{center}
\mbox{\epsfxsize=8.2cm\epsffile{lambda_B_mu_0.00_n_1_LogLog.eps}}  \\
\vspace{1.cm}
\mbox{\epsfxsize=8.cm\epsffile{lambda_B_mu_0.00_n_LogLog_0.001_0.5.eps}}  
\end{center}
\caption{(Color online) Spectrum of the W-C model as a function of the
  coupling constant $\lambda(B)$ in the low energy limit. Upper panel
  corresponds to n=1 states with different values of $\kappa=0,1,..$ .
  The case $\kappa=0$ is compared
to the non relativistic solution. Horizontal lines correspond respectively 
to  the maximal values of the coupling constant  for a well-defined
ground state  ($\lambda=2$)   and  to the minimal value   for
which  the abnormal solutions exist ($\lambda=1/4$).
Lower panel contains the full spectrum for $n\le6$ and $\kappa\le3$
to make explicit the different crossings. Notice that the unphysical
solutions with odd $\kappa$ (giving no contribution to the S-matrix)
are naturally inserted in the spectrum.
\label{lambda_B}}
\end{figure}

These different behaviours are illustrated in Fig. \ref{lambda_B} where
we have displayed the dependence of the coupling constant
$\lambda={\alpha\over\pi}$ on the binding energy $B$
for the lowest solutions of the W-C model. 
Upper panel contains only the n=1 states with $\kappa=0,1,2,3,4$. 
The curve corresponding to $\kappa=0$ (black solid  line) is tangent
to the non-relativistic  (NR) one  (black dashed line) from which
it departures logarithmically, as it is visible, starting at
$B\approx 0.001$.
The perturbative results, provided by Eq. (\ref{B_Pert}), are
indistinguishable from the exact ones in the  considered energy range.
Those corresponding to  $\kappa>0$ (colored solid lines)  do not have any
non-relativistic counterparts. 
Lower panel represents the spectrum for $n\ge 1$ states and different
values of $\kappa=0,1,2,3$.
The horizontal line  $\lambda=2$ ($\alpha=2\pi$) indicates  the maximal
value of the coupling constant ensuring a well-defined ground state. 

The normal and abnormal solutions have also different domains of
existence with respect to the coupling constant $\alpha$. 
As a mathematical solution of the BS equation (\ref{bs}), the normal
solutions exist for any (positive) values of $\alpha$, although, as we
have already discussed,
they have a clear physical meaning only in the range $\alpha\in[0,2\pi]$,
where $M^2\ge 0$.
However, the very existence of the abnormal solutions (all of them)
requires a coupling constant greater than some critical value,
$\alpha\ge{\pi\over4}$ ($\lambda\ge1/4$). 
This can be clearly seen from the results of Fig.  \ref{lambda_B}, where
all the abnormal states (color line) were
found above the horizontal $\lambda=1/4$ line. 

Wick and Cutkosky \cite{wick,cutk} found the following approximate
analytic expression for the abnormal spectrum near the continuum
threshold ($B\to 0$):  
\begin{equation} \label{abnspectr}
  M_{n\kappa}^2\simeq 4m^2\left(   1 - e^{    -{(\kappa-1)\pi/
   \sqrt{{\alpha\over \pi}-{1\over 4}}}}\right),\;\kappa=2,3,\ldots\ ,      
\end{equation}
where the condition $\alpha>\pi/4$   is explicitly obtained. 
At $B/m\ll 1$ this spectrum vs. $\kappa$ does not depend on $n$.
It also indicates that to be able to distinguish abnormal states from the
continuum threshold on experimental grounds, the coupling constant
should be increased at least up to values  of $\alpha\approx 4\div 5$;
otherwise, for values of $\alpha$ very close to $\pi/4$, the
exponential in Eq. (\ref{abnspectr}) is nearly zero and the discrete
spectrum becomes hardly distinguishable from the continuum.

It is worth noting that the existence of a lower bound of the
coupling constant for the abnormal solutions is reminiscent of the
massive-exchange case, i.e. $\mu\ne0$ in the kernel  (\ref{ladder}),
which was considered with some detail in \cite{MC_PLB474_2000} both in
the BS and the Light-Front Dynamics frameworks.
The $B_{\mu}(\alpha)$ dependences (Figs. 5 and 7 of
\cite{MC_PLB474_2000}) are similar to those displayed in Fig.
\ref{lambda_B}), what suggest the possibility 
to associate a mass with the abnormal states.
However, essential differences in the number of bound states for a
given $\mu$ remain: infinite in the W-C model and (at most) finite
in the massive case.

In summary,  the range of the coupling constants to be considered in
this model is  $0<\alpha<2\pi$  ($0<\lambda<2$) for the
normal states, i.e. $\kappa=0$, and ${\pi\over 4}<\alpha<2\pi$
($ 1/4<\lambda<2$) for the abnormal ones ($\kappa>0$).
On another hand,  according to Refs. \cite{cia,nai}, the abnormal
solutions with odd
values of $\kappa$ do not contribute to  the $S$-matrix and therefore only
those with even $\kappa$ can have a physical meaning. 
In the subsequent part of this work, we will concentrate on the latter case.
Furthermore we will  restrict ourselves,  to the $l=0$ states with $n=1$
and $n=2$.

For  $n=1$ states, the sum (\ref{Phi}) is reduced to a single term
involving only the function $g_1^0$ satisfying the homogeneous integral
equation
\begin{equation}\label{g10_Int}
  g_1^0(z)=\frac{\alpha}{2\pi}\int_{-1}^1 \frac{R(z,z')}{Q(z')} \,
  g_1^{0}(z')dz',
\end{equation}
or, equivalently, in its differential form
\begin{equation}\label{g0}
{g''}_1^0(z)+\frac{\alpha}{\pi}\frac{1}{(1-z^2)Q(z)}g_1^0(z)=0,
\end{equation}
with the boundary conditions $g_1^0(\pm 1)=0$.
The corresponding BS amplitude is expressed in terms of $g_1^0$ as
\begin{equation}\label{Phi10}
  \Phi_1(k,p)=\int_{-1}^1\frac{-im^3 \,g_{1}^0(z)dz}{[m^2-\frac{1}
      {4}M^2 -k^2-p\cd k\,z-\imath\epsilon]^{3}}.
\end{equation}

\bigskip
For  $n=2$ states, the sum (\ref{Phi}) involves  two functions $g_2^0$
and $g_2^1$. The function $g_2^0$ satisfies the homogeneous integral
equation:
\begin{equation}\label{g20_Int}
  g_2^0(z)=\frac{\alpha}{4\pi}\int_{-1}^1 \frac{R^2(z,z')}{Q(z')} \,
  g_2^{0}(z')dz',
\end{equation}
and in differential form
\begin{eqnarray}\label{eq1c}
{g''_2}^0(z)&+&\frac{2z}{(1-z^2)}{g'}_2^0(z)-\frac{2}{(1-z^2)}g_2^0(z)
\nonumber\\
&+&\frac{\alpha}{\pi}\frac{1}{(1-z^2)Q(z)}g_2^0(z)=0,
\end{eqnarray}
with the boundary conditions $g_2^0(\pm 1)=0$, while $g_2^1$ is determined
from $g_2^0$ through the integral equation
\begin{eqnarray}\label{g21}
  g_2^1(z)&=&   \frac{\alpha}{6\pi}\int_{-1}^1\frac{R(z,z') }
  {[Q(z')]^2}
  \,g_2^{0}(z')dz'    
  \nonumber\\
  &+&   \frac{\alpha}{2\pi}\int_{-1}^1\frac{R(z,z')}
     {Q(z')}  \, g_2^{1}(z')dz',
\end{eqnarray}
which can also be rewritten in the form of an inhomogeneous
differential equation:
\begin{eqnarray}\label{eq21c}
&&  {g''}_2^1(z)+\frac{\alpha}{\pi}\frac{1}{(1-z^2)Q(z)}g_2^1(z)
\nonumber \\
&&\ \ \ \ \ \ \ \ \ \ \ \   
= - \frac{\alpha}{3\pi}\frac{1}{(1-z^2)[Q(z)]^2}g_2^0(z).
\end{eqnarray}
\par
The BS amplitude (\ref{Phi}) is now expressed in terms of two functions
$g_2^1$ and  $g_2^0$:
\begin{eqnarray}\label{Phi2}
  \Phi_2(k,p)&=&\int_{-1}^1\frac{-im^3 \,g_{2}^1(z)dz}
      {[m^2-\frac{1}{4}M^2 -k^2-p\cd k\,z-\imath\epsilon]^{3}}
      \nonumber\\
      &+&\int_{-1}^1\frac{-i m^5\,g_{2}^0(z)dz}
      {[m^2-\frac{1}{4}M^2 -k^2-p\cd k\,z-\imath\epsilon]^{4}}.
\end{eqnarray}

\subsection{Numerical solutions for some selected states $g_{n\kappa}^k$}\label{g_num}

We present, in this subsection, the numerical results concerning the
first  states of the W-C spectrum. 
We fix hereafter $m=1$ and  the coupling constant to the value $\alpha=5$.
We will consider along the work an ensemble of states with $n=1,2$ and
$\kappa=0,2,4$ that,
for the sake of simplicity in notation, will be numbered with No. 1-6
in the Tables \ref{tab1} and \ref{tab2}.

\begin{table}[h!]
\begin{center}
\caption{Binding energy $B$ and two-body norm ($N_2$)
of the low-lying normal  ($\kappa=0$) and abnormal
($\kappa=2$) states, for the coupling
constant value $\alpha=5$.}\label{tab1}
\begin{tabular}{cccll}
\hline\noalign{\smallskip} 
No.&   n & $\kappa$ &  $B$ & $N_2$ \\
\noalign{\smallskip}\hline 
\noalign{\smallskip}
1&1&0  &   0.999259         &0.65 \\
2&2&0  &   0.208410         & 0.61 \\
3& 1 &2  &
$3.51169 \cdot 10^{-3}$
& 0.094 \\
4& 2 &2 &
$1.12118 \cdot 10^{-3}$
& 0.077 \\
\noalign{\smallskip}\hline
\end{tabular}
\end{center}
\end{table}

\begin{figure}[h!]
\vspace{.8cm}
\begin{minipage}[h!]{8.7cm}
\begin{center}
\epsfxsize=7.cm\epsfysize=5.cm\epsfbox{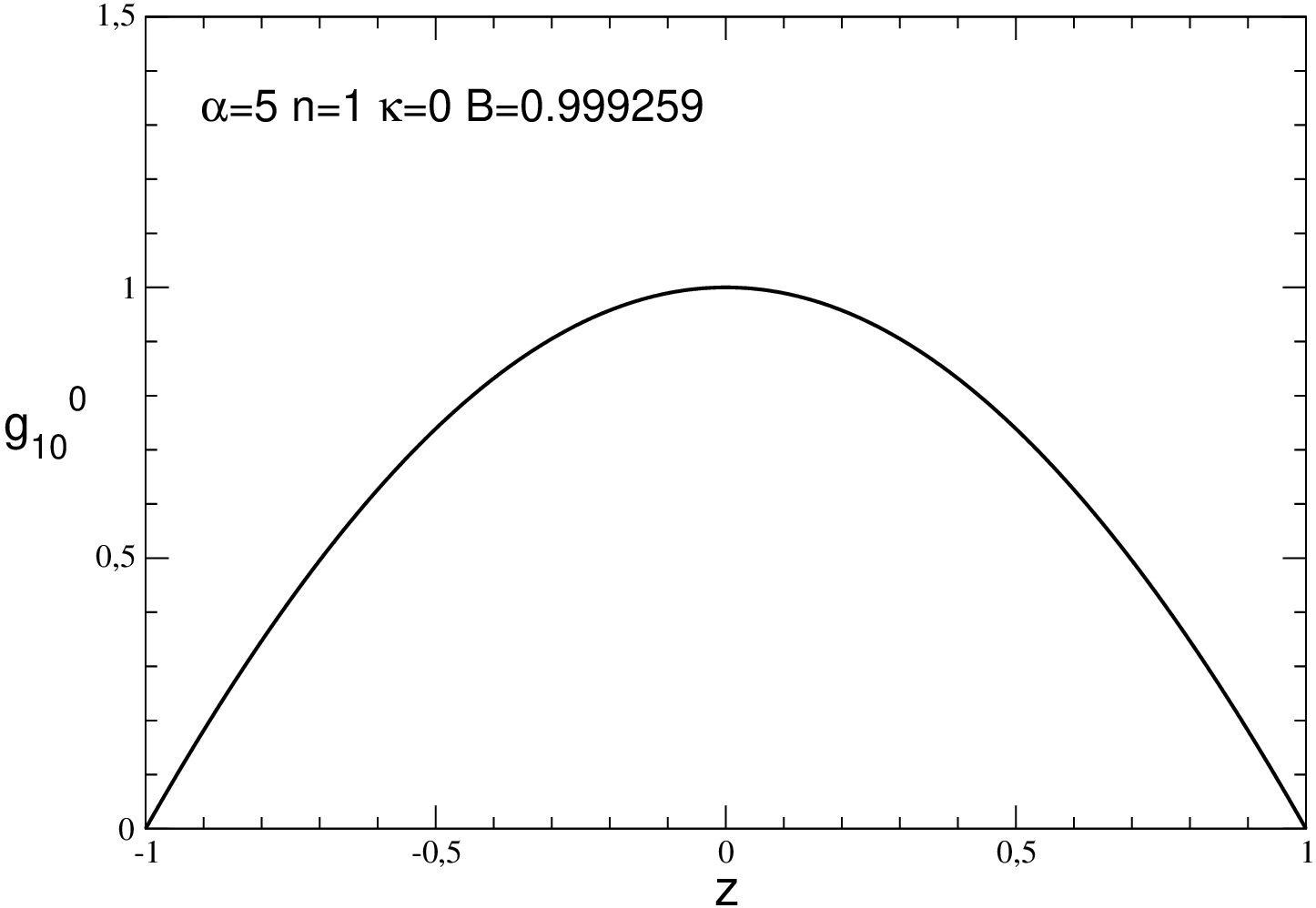}
\caption{$g_{10}^0$ for the  normal state No. 1 of Table \ref{tab1}.} \label{fig1}
\end{center}
\end{minipage}

\begin{minipage}[h!]{8.7cm}
\vspace{.9cm}
\begin{center}
\epsfxsize=7.cm\epsfysize=5.cm\epsfbox{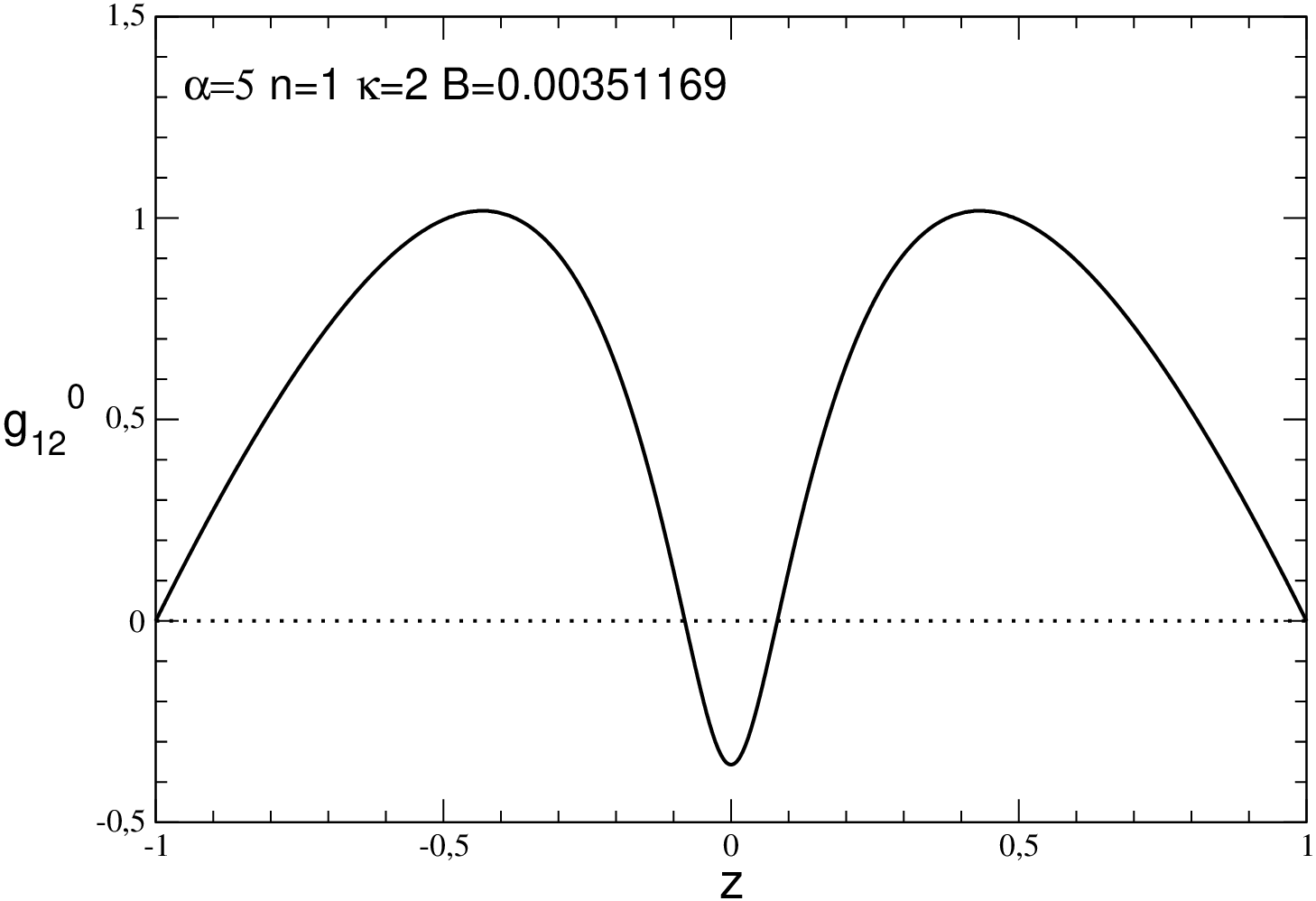}
\caption{$g_{12}^0$  for the   abnormal state No. 3  of Table \ref{tab1}.}\label{fig1p}
\end{center}
\end{minipage}
\end{figure}

The  binding energies for the lowest $n=1,2$  normal ($\kappa=0$) and
abnormal ($\kappa=2$) states  $g_{n\kappa}^{\nu}$ are presented in Table
\ref{tab1}.
All $\nu=0$ components are arbitrarily normalized to $g_{n\kappa}^0(0)=1$. 
The corresponding solutions for the $n=1$ states -- $g_{10}^0$  and
$g_{12}^0$  -- are displayed in  Figs. \ref{fig1} and \ref{fig1p}.
They have comparable sizes
and their nodal structure is determined by $\kappa$ only.

\begin{figure}[h!]
\vspace{0.8cm}
\begin{center}
\epsfxsize=7cm\epsfysize=5cm\epsfbox{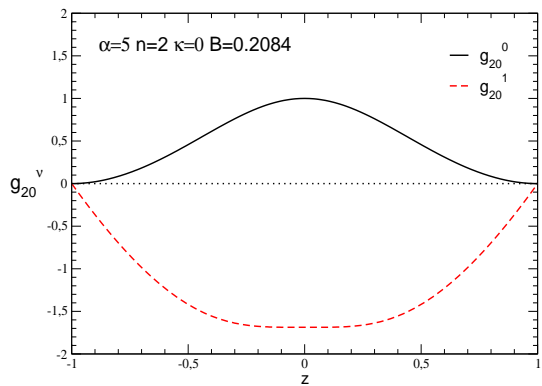}
\caption{Components $g_{20}^0$ and $g_{20}^1$  of  the normal state No. 2 from Table \ref{tab1}.}\label{g2}
\end{center}
\end{figure}

\begin{figure}
\vspace{0.8cm}
\begin{center}
\epsfxsize=7cm\epsfysize=5cm\epsfbox{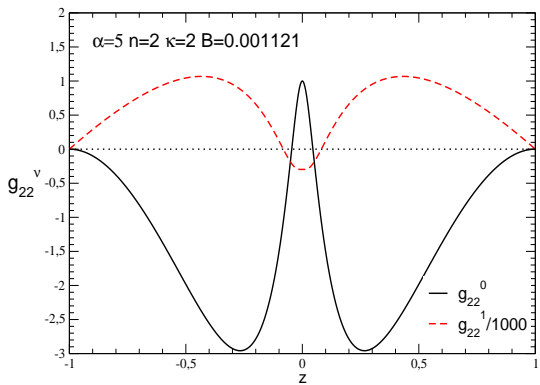}
\caption{$g_{22}^0$ and  $g_{22}^1$  (scaled by a factor $10^3$) of the abnormal state No. 4 (Table \ref{tab1}).}
\label{g2ab}
\end{center}
\end{figure}

The two components $g_{2\kappa}^0$ and $g_{2\kappa}^1$ of the $n=2$
states    are  plotted in  Fig. \ref{g2}   (state No. 2 with $\kappa=0$)
and Fig. \ref{g2ab} (state No. 4 with $\kappa=0$). 
The component $\nu=1$ is dominant in both cases, but for the state No. 4
it is $\sim 1000$ times larger (see the scaling factor in Fig. \ref{g2ab}).
This enhancement is due to the $Q^2$ factor in the denominator of the
right-hand-side of Eq. (\ref{eq21c}), 
which, in the limit $B\to 0$ and around $z=0$, behaves as
$Q^2\approx B^2 $.
Thus, for  $n>1$ states with small binding energies, the component
$g_{n\kappa}^0$ that determines $M^2$ is
negligibly small with respect to the other components.
We will see, however, in the following section that they all
play equivalent roles in the construction of the BS amplitude itself 
and in the form factors.

\vspace{0.5cm}

\vspace{0.5 cm}
In the rightest column of Table \ref{tab1} we have  also included the
norm $N_2$ of the two-body contributions in the Fock space, as it is defined in   \ref{appN2}.
For the states with $n=1$, $N_2$ is given by Eq. 
of \ref{appN2}
and for $n=2$, by Eqs. ({\ref{norm2}). We remark  therein that the
two-body norm $N_2$ for the abnormal ($\kappa=2$) states  is much smaller
than for the normal ($\kappa=0$) ones.
This comparison concerns however  states covering the two extreme cases
in the spectrum:  deeply bound states (Nos. 1 and 2) and
nearthreshold ones (Nos. 3 and 4).
To better understand this difference,  we have 
studied the dependence of $N_2$ on the binding energy for the first
normal and abnormal states. Results are displayed in  Fig. \ref{N2_B}.

\begin{figure}[h!]
\vspace{0.9cm}
\begin{center}
\epsfxsize=7. cm\epsfysize=5cm\epsfbox{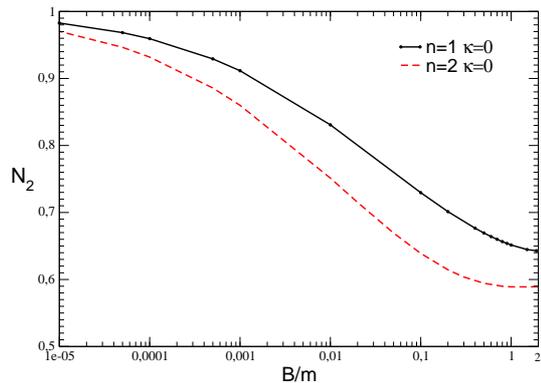}\\\vspace{1.cm}
\epsfxsize=7. cm\epsfysize=5cm\epsfbox{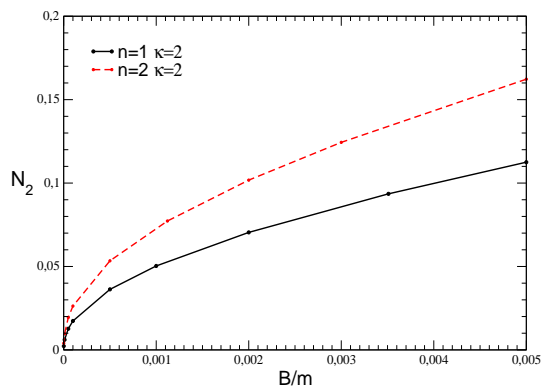}
\caption{(Color online) $N_2$-dependence  on the binding energy for
  the  $n=1$ and $n=2$ states: normal (upper panel) and abnormal (lower
  panel).}
\label{N2_B}
\end{center}
\end{figure}

The upper panel concerns the normal states. 
The behaviours of the the $n=1$  (black solid line) and $n=2$  (red
solid line) states are quite similar:  
$N_2$ decreases monotonically from $N_2\approx 1$ when $B\approx 0$
down to an asymptotic value  when $B\to 2m$. We found numerically
$N_2(2m)\approx 0.64$  for n=1 and $N_2(2m)\approx 0.59$ for n=2.
For the ground state n=1, these limiting values  were found analytically
in \cite{dshvk}, as well as  the perturbative expansion
in their vicinity. Thus, the limit $B\to0$ is described by Eq.
(\ref{N2_1a}), i.e., 
\[ N_2(B)= 1+{1\over\pi}\sqrt{{4B\over m}}\ln\left({4B\over m}\right). \]
At $B/m=10^{-5}$, this perturbative expansion gives $N_2=0.980$ in
close agreement with  the black curve of Fig. \ref{N2_B}.
We conclude from this study that the normal states are dominated by
two-body norms. This is particularly true in the  limit $B\to 0$,
where $N_2\to 1$, but remains also true  in all the energy domain,
although decreasing with increasing $B$.

A very different behaviour is observed with the abnormal states,
represented in the lower panel of  Fig. \ref{N2_B}.
As one can see, the two body norm $N_2$ of these states not only remains
comparatively very small, but also vanishes in the non-relativistic
limit, making them, in this region, genuine many-body states.
The one order of magnitude observed in Table \ref{tab1} for the
binding energies hides in fact
a deeper and striking difference between normal and abnormal BS states,
independent of their comparison with the non-relativistic spectrum.
It is provided by their two-body content: abnormal states do not have
in the limit $B\to 0$ any two-body contribution and have, thus,
genuine many-body structures.
Beyond this limit the norm of the two-body sector remains extremely
small. This is the reason why they are absent in the non-relativistic
limit reduced to the two-body Schr\"odinger equation.

The results presented in Table \ref{tab1}    are  completed in
Table \ref{tab2}  by studying the $\kappa=4$  excitations of $n=1,2$ 
states. The same conclusion holds, even in a more dramatic way.
Their two body norms are one order of magnitude smaller than for the
$\kappa=2$ states of Table  \ref{tab1}.
This can be expected due to their smaller binding energies and in view
of the behaviour described in the lower panel of Fig. \ref{N2_B}.

\begin{table}[h!]
\begin{center}
  \caption{Same as in Table \ref{tab1}, for the abnormal states with   $\kappa=4$.}  \label{tab2}
\begin{tabular}{cccll}
\hline\noalign{\smallskip} 
No.&   n & $\kappa$ &  $B$ & $N_2$ \\
\noalign{\smallskip} \hline
\noalign{\smallskip}
5&1  &4&
$1.54091\cdot 10^{-5}$ & $6.19\cdot 10^{-3}$ \\
6&2  &4&
$4.95065    \cdot 10^{-6}$ &
$2.06\cdot 10^{-5}$\\
\noalign{\smallskip}\hline
\end{tabular}
\end{center}
\end{table}

The abnormal solution $g_{14}^0$} for $n=1, \kappa=4$  state (No. 5
in Table \ref{tab2}), is shown in Fig. ~\ref{fig10_14}, displaying its
more involved nodal structure (4 zeros in  $]-1,+1[$). 
The functions {$g_{24}^{\nu}$ of  the abnormal state  $n=2, \kappa=4$
(No. 6 in Table \ref{tab2}), are plotted  in Fig. \ref{g240_g241_Nb6}.
The extreme smallness of the binding energy of this state
generates a huge enhancement factors in the inhomogeneous equation
(\ref{eq21c}) (through the factor Q) which results in 
a huge  dominance of the $\nu=1$ component in the full BS amplitude.
Notice that  $g_{24}^1$ has been reduced by a factor $10^5$ to become
comparable with $g_{24}^0$.

\begin{figure}[h!]
\vspace{1.2cm}
\begin{minipage}[h!]{8.5cm}
\begin{center}
\epsfxsize=8.cm\epsfysize=5cm\epsfbox{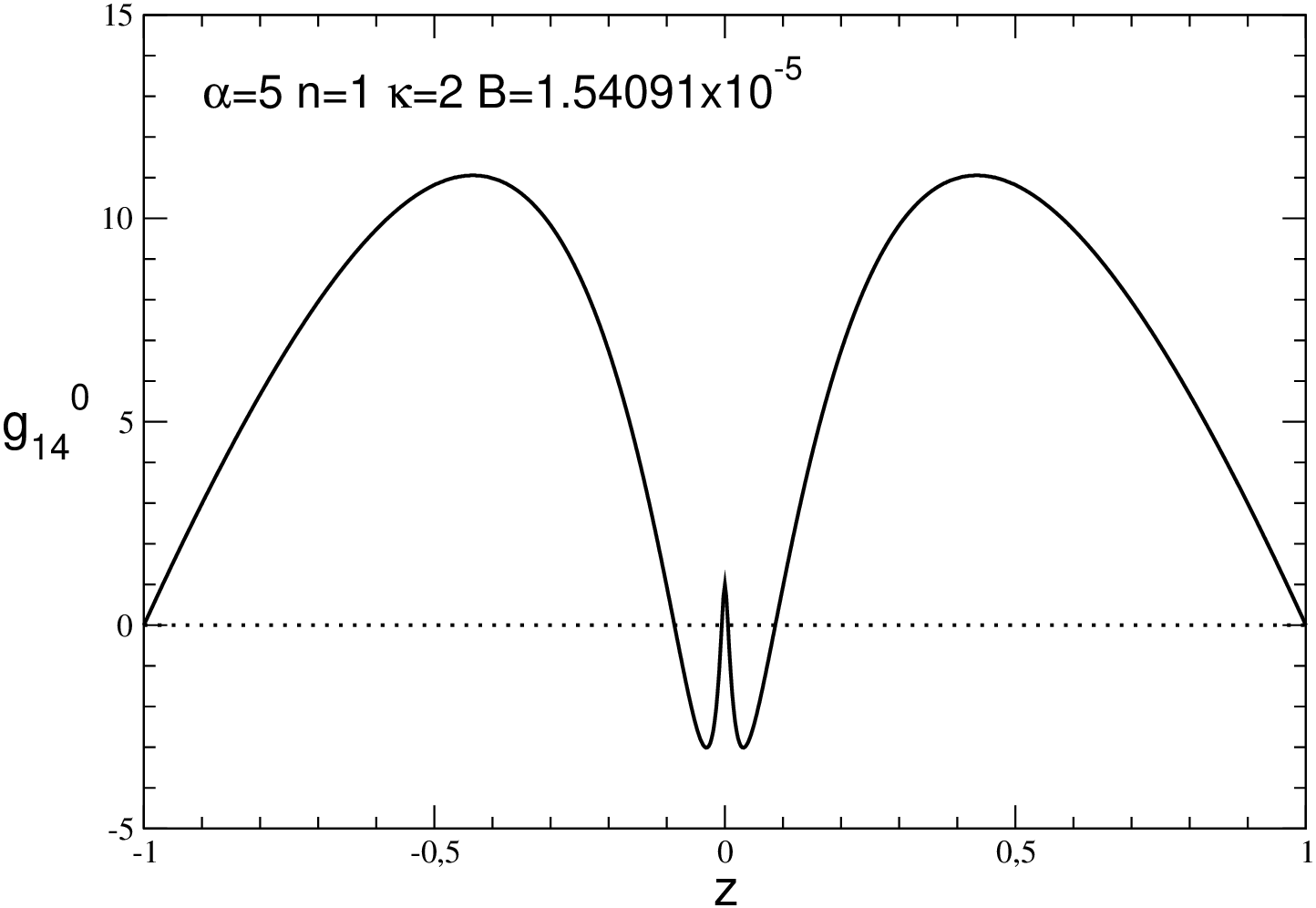}
\caption{$g_{14}^0$ from state No. 5 in Table \ref{tab2}.}
\label{fig10_14}
\end{center}
\end{minipage}
\vspace{1cm}\\
\begin{minipage}[h!]{8.5cm}
\begin{center}
\epsfxsize=8.cm\epsfysize=5cm\epsfbox{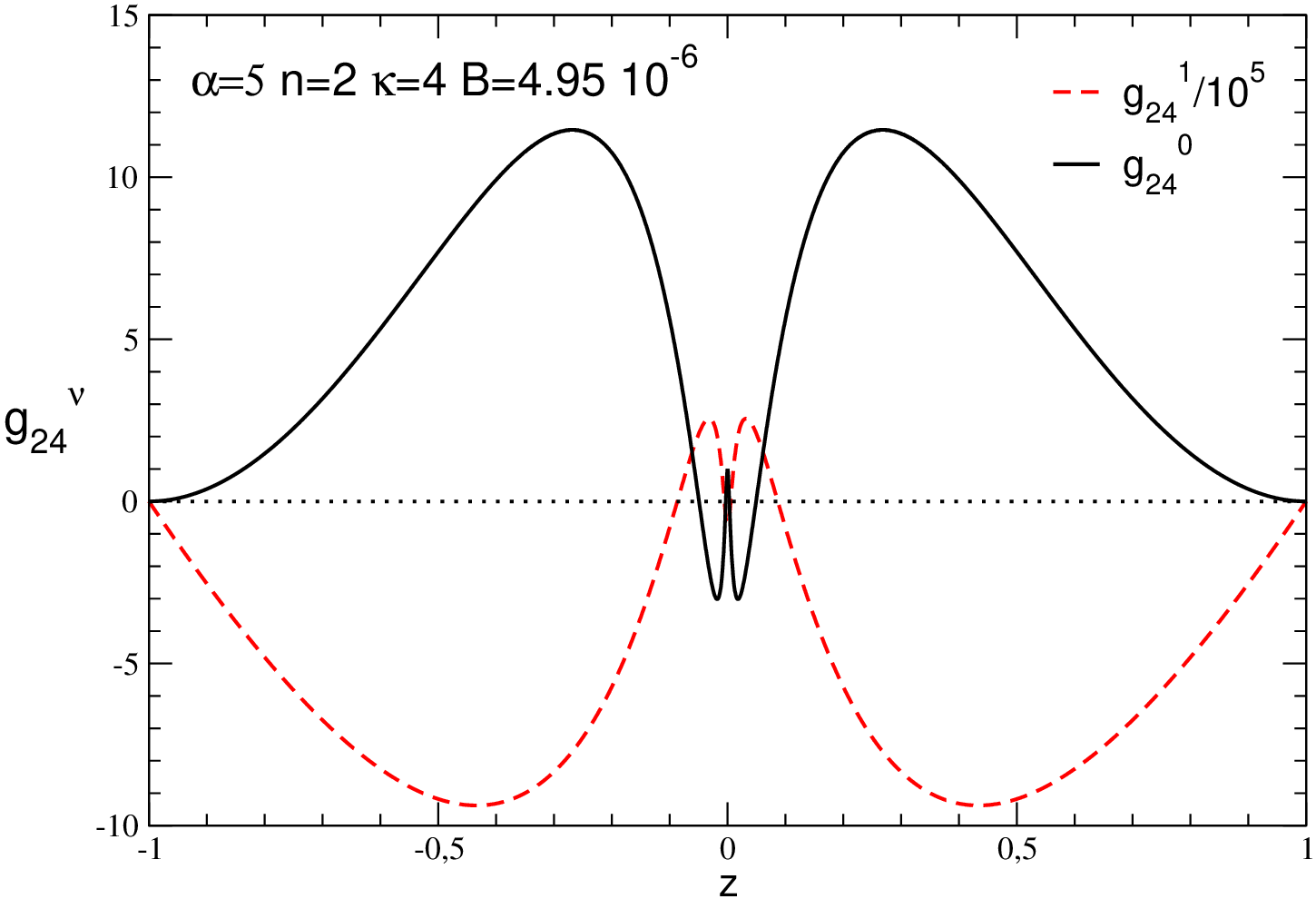}
\caption{(Color online) $g_{24}^{\nu}$ from state No. 6 in Table \ref{tab2}.}
\label{g240_g241_Nb6}
\end{center}
\end{minipage}
\end{figure}

\section{Electromagnetic form factors}\label{FFs}

We suppose that one of the two constituent particles is charged. 
The electromagnetic form factor of the system can be expressed in
terms of its BS amplitude. It is enough to consider  inelastic
transitions from an initial $| i\rangle$ to a final $| f\rangle$ state.
The  elastic  form factors are obtained from them as a particular case,
with $f=i$.
\begin{figure}[h!]
\centering
\includegraphics{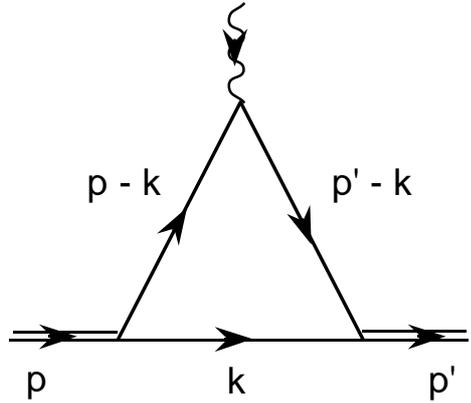}
\caption[*]{Feynman diagram for the electromagnetic form factor.
\label{triangle}}
\end{figure}
\par
The electromagnetic vertex $J_{\mu}$, corresponding to a  transition
$| i\rangle\to | f\rangle$ is shown graphically in Fig.~\ref{triangle}. 
The corresponding vertex amplitude reads (we use Itzykson and Zuber
\cite{IZ} conventions for the Feynman rules):
\begin{eqnarray}\label{ffGam}
iJ_{\mu}&=&\int \frac{d^4k}{(2\pi)^4}\,
\frac{i[-i(p+p'-2k)_{\mu}]}{(k^2-m^2+i\epsilon)}
\nonumber\\
&\times&\frac{i\left[-i\overline{\Gamma}
\left(\frac{1}{2}p' -k,p'\right)\right]}
{[(p'-k)^2-m^2+i\epsilon]}
\frac{i\left[i\Gamma \left(\frac{1}{2}p-k,p\right)\right]}
{[(p-k)^2-m^2+i\epsilon]},
\end{eqnarray}
where $\Gamma(k,p)$ is the vertex function, related to
the BS amplitude by the equation
\begin{eqnarray}\label{PhiG}
& &\Phi(k,p)=\frac{\Gamma(k,p)}
{\left[(\frac{p}{2}+k)^2-m^2+i\epsilon\right]
\left[(\frac{p}{2}-k)^2-m^2+i\epsilon\right]}.\nonumber \\
& &
\end{eqnarray}
$\overline{\Gamma}$ is the conjugate of $\Gamma$, obtained from the
latter by complex conjugation and use of the anti\-chrono\-logical
product. A similar definition also holds for the BS amplitude
$\overline{\Phi}$
\cite{nak69}.
\par
The electromagnetic vertex for the transition $i\to f$ is expressed in
terms of the BS amplitude as
(see e.g. Eq. (7.1) in \cite{cdkm}):
\begin{eqnarray}\label{ffbs}
J_{\mu}
&=&i\int \frac{d^4k}{(2\pi)^4}(p+p'-2k)_\mu \; (k^2-m^2)
\nonumber\\
&\times&\overline{\Phi}_f\left(\frac{1}{2}p'-k,p'\right)\Phi_i
\left(\frac{1}{2}p-k,p\right). 
\end{eqnarray}
It has the following general decomposition in terms of two scalar
functions (see\footnote{We change the notations in comparison to Ref.
\cite{tff}, where the form factor $G$ was denoted $F'$.} \cite{tff}):
$F(Q^2)$ and $G(Q^2)$ 
\begin{eqnarray}\label{ffc}
J_{\mu}&=&\left[(p_{\mu}+{p'}_{\mu})+ ({p'}_{\mu}-p_{\mu})\frac{Q_c^2}
{Q^2}\right]F(Q^2)
\nonumber\\
&-& ({p'}_{\mu}-p_{\mu})\frac{Q_c^2}{Q^2}G(Q^2).
\end{eqnarray}
Here, $q=p'-p$, $Q^2=-q^2=-(p'-p)^2$ and
\begin{equation} \label{Qc2}
Q_c^2={M_f}^2-M_i^2,
\end{equation}  
$M_i$ and $M_f$ being the masses of the initial and final states,
respectively.

Since $q\cd J=Q_c^2 G(Q^2)$, the above decomposition does not suppose,
in general, current conservation $q\cd J=0$, which implies 
\begin{equation}\label{G}
G(Q^2) \equiv 0.
\end{equation} 
A direct proof of this result from the BS equation is presented in
 \ref{proofG}.
The current conservation becomes a stringent self-consistency criterion
of our results: the calculated form factor $G(Q^2)$ should
be identically zero, or very small  within the numerical uncertainties.
 
We also note that since $J_{\mu}$ [Eq. (\ref{ffc})] is not singular
at $Q^2=0$, the following  relation should hold:
\begin{equation}\label{FeqFp}
F(0)=G(0). 
\end{equation}
Equation (\ref{G}) then implies that $F(0)=0$ for the transition form
factors (for which $Q_c^2\neq 0$).
\par
From Eq. (\ref{ffc}), the form factors are expressed throu\-gh
$J_{\mu}$ as
\begin{eqnarray}\label{FFp}
& &F(Q^2)  = \frac{ (p+p')\cd J\,Q^2 + q\cd J \, Q_c^2}
{[(M_f-M_i)^2 +Q^2][(M_f+M_i)^2+Q^2]},\nonumber \\
& &G(Q^2) = \frac{q\cd J}{Q_c^2}.
\end{eqnarray}
Considering these formulas at $Q^2=0$ (which does not imply $q=0$),
one gets the relation
$F(0)=\left.\frac{q\cd J}{Q_c^2}\right|_{Q^2=0}=G(0)$, which
reproduces Eq. (\ref{FeqFp}). 
\par
The expressions for the form factors are obtained by substituting
in Eqs. (\ref{FFp}) the current $J$ [Eq. (\ref{ffbs})], then
substituting the BS amplitudes (\ref{Phi}), using the Feynman
parametrization and integrating over $k$. In this way, we find the
form factors in the form of integrals over products of functions
$g_n^{\nu}(z)$ and $g_{n'}^{\nu'}(z')$. (Details of similar calculations can
be found in Ref. \cite{ckm_ejpa}.) 
The result for the transition form factor $F(Q^2)$ can be written
in the form:
\begin{equation}\label{ff}
F(Q^2)=\sum_{\nu=0}^{n-1}\sum_{\nu'=0}^{n'-1}F_{nn'}^{\nu\nu'},
\end{equation}
and similarly for $G(Q^2)$.
The expressions of the functions of the right-hand-side of this
equation for the cases $n=n'=1$ and $n=n'=2$ are given in
\ref{exprssff}.

In the following subsections, we examine the numerical results for the
elastic and transition form factors for some states from Tables
\ref{tab1} and  \ref{tab2}, corresponding to the coupling constant
$\alpha=5$.

\subsection{Elastic form factors $F_e(Q^2)$}\label{el}

There are two regions of interest in studying the elastic form factors
($F_e$):
the region near the origin, giving insight into the size of the system,
and the asymptotic region $Q^2\gg m^2$, related to the many-body structure
of the wave function \cite{matvmurtavk,brodsfarr,radyush}.

With the elastic form factor of a bound state, normalized to
$F_e(Q^2=0)=1$, the squared radius is given by
\begin{equation}\label{r2}
<r^2>= -6 \left(dF_e\over dQ^2\right)_{Q^2=0} 
\end{equation}
and the root mean squared (r.m.s.) radius by $R=\sqrt{<r^2>}$. 
In the non-relativistic theory, the size of a  bound state scales, as
a function of its binding energy, as $R\approx {1\over\sqrt{mB}}$. 

On the other hand, as mentioned in the Introduction, the asymptotics
of $F_e$ should qualitatively probe the compositeness of the state. 
According to \cite{matvmurtavk,brodsfarr,radyush},  the elastic form
factors of a $n$-body system should decrease as $1/(Q^2)^{n-1}$,
where $n$ is the number of the constituents of the state.
It is, however, worth emphasizing here some essential differences of
the W-C model with the theoretical framework in which the above
asymptotic behaviors have been obtained. The latter have been derived
in theories characterized by dimensionless coupling constants, like
QCD and the parton model. In the W-C model, bosonic fields interact
by the exchange of a scalar particle; the coupling constant $g$ is
then dimensionful, having the dimension of mass.
This has an immediate consequence on the behavior of the BS
amplitude at large momenta. In the W-C model, as can be checked from
Eqs. (\ref{bs}) and (\ref{ladder}), the BS amplitude behaves at large
momenta as $1/(k^2)^3$. In QCD, in the ladder approximation, where
quarks interact by means of an exchange of a gluon field, the
scalar part of the BS amplitude behaves at large momenta as
$1/(k^2)^2$, up to logarithms. Extending the comparison to bosonic
$\phi^4$-like theories, where the coupling constant is also dimensionless,
the interaction between two bosonic constituents is realized either
by contact terms, or by the exchange of a two-particle loop; in both
cases the large-momentum behavior of the BS amplitude is again 
$\frac{1}{(k^2)^2}$, up to logarithms. The faster decrease of the
BS amplitude in the W-C model affects the behaviors of the form factors: 
one thus expects in this model behaviors of the type $1/(Q^2)^n$, up
to logarithms, instead of $1/(Q^2)^{n-1}$. 

In QCD, the number $n$ represents the number of valence quarks
(including, eventually, the number of valence gluons, in the case
of hybrids).
The Fock sectors of the state which contain the sea quarks, therefore
with higher $n$'s, are expected to decrease asymptotically faster and
thus to display rapidly the asymptotic dominance of the valence quark
sector, a phenomenon well observed on experimental grounds. In the
W-C model, for the normal solutions, one indeed expects the dominance
of the two-body sector, as was concluded in Sec. \ref{WCsol}.
However, for the abnormal solutions, the two-body sector is weakly
contributing to the composition of the corresponding states, which are
dominated by higher sectors of the Fock space. One therefore expects
here a competition between the contributions of the higher sectors
of the Fock space, which asymptotically decrease more rapidly, but
have large coefficients, and the contribution of the two-body sector,
which dominates in the asymptotic region, but with a small coefficient.

For abnormal solutions, or hybrid states, of the \mbox{W-C} model, a 
refined analysis necessitates the distinction between three regimes in
the $Q^2$ evolution of the form factors, rather than two:
the very small $Q^2$ regime, determined by the binding energy (and 
equivalently by the rms radius), 
the intermediate $Q^2$ region where the decrease is determined by
the many-body components (and therefore is fast)
and  the asymptotic region, where the many-body contribution is
exhausted, and only  the two-body contribution survives. 
Since in the W-C model the content of any state is a two-body component
plus an indefinite number of  exchange particles, 
the asymptotic behavior of all the elastic form factors is finally
determined by its two-body contribution. 
Therefore, the asymptotic $Q^2$-dependence should be the same, though
with different coefficients, for all elastic form factors.  
In particular, we predict that the ratios of all elastic form factors
should tend to constants at $Q^2\to\infty$.

We first examine the $n=1$ states. They are defined by a single component
$g_{1\kappa}^0$, the same that determines the binding energy.
We have plotted, in Fig. \ref{Fel_1_3}, in solid lines, the elastic
form factors for the two $n=1$ states of Table \ref{tab1}:  on top,
the normal state No. 1 ($\kappa=0$ , $B=0.999$), and at bottom, the
abnormal state No. 3 ($\kappa=2$, $B=0.00359$). 

\vspace{0.5cm}
\begin{figure}[h!]
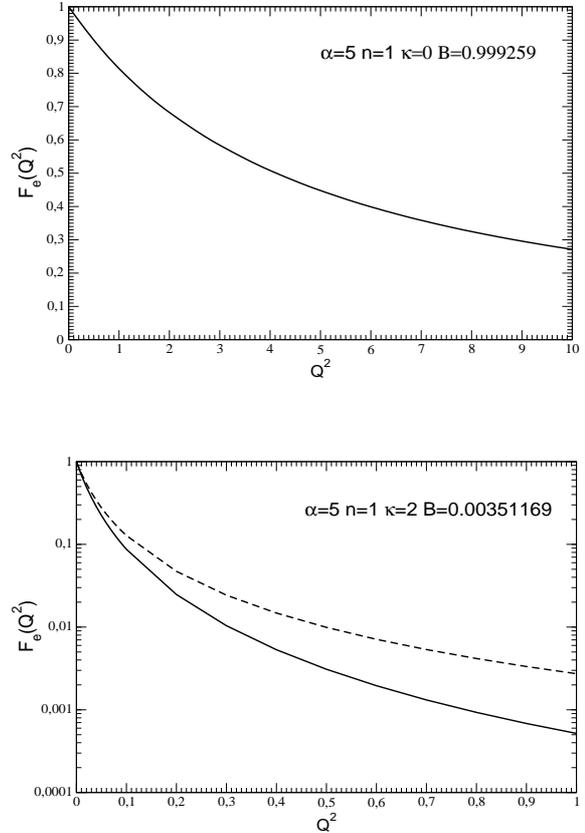

\begin{center}
\epsfxsize=7.5cm\epsfysize=5cm\epsfbox{FQ2_Nb1.eps}\vspace{1cm}\\
\epsfxsize=7.5cm\epsfysize=5cm\epsfbox{FQ2_Nb3.eps}
\end{center}
\caption{Elastic form factors of $n=1$ states of Table \ref{tab1}
(solid lines): No. 1 (normal) on top  and  No. 3 (abnormal) at bottom;
the dashed line corresponds to the state with $\kappa=0$  with the same
binding energy (and rms radius) as the state No. 3 with
$\kappa=2$.}
\label{Fel_1_3}
\end{figure}

The corresponding rms radii are $R_1=1.16$ fm and $R_3=15.7$ fm,
respectively, which roughly scale as $\sim{1\over \sqrt{mB}}$.
Both states have only one component $g_{n\kappa}^0$, which in turn
determines the energy. It is then natural that they have a similar
behaviour, close to that of the non-relativistic one.
At this level, one cannot see any drastic difference between a normal
and an abnormal state. 
The behaviours of their form factors are quite similar, however with
a much faster decrease
for the state No. 3, as expected from its smaller binding energy
and its many-body structure. At $Q^2=1$, the value of $F_e$ for the
state No. 3 is three orders of magnitude smaller than that of the
state No. 1.
In order to disentangle the contributions of the binding
energy and of the asymptotic behaviour,
we have adjusted, at a second step, $\alpha$ of the state No. 1 to
have the same binding energy as the state No. 3. 
The result is displayed in dashed line on the lower panel: both curves
are tangent to each other at the origin, implying that the rms radii
are the same, but one notices that the abnormal state form factor still
decreases much faster than that of the normal one, by a factor 10 at
$Q^2=1$. 

\newcommand\egal{\mathop{=}}
One can show that the elastic form factors behave, when
$Q^2/m^2\to \infty$ as
\begin{eqnarray}\label{Fe_asymp}  
  F_e(Q^2)&=&\Big({m^2\over Q^2}\Big)^2
  \left[c_2  \ln \left({Q^2\over m^2}\right) + c_0 \right] \nonumber\\
  &+& {\mathcal O}\left(\Big({m^2\over Q^2}\Big)^3\ln \left({Q^2\over m^2}\right),
  \Big({m^2\over Q^2}\Big)^3\right).
\end{eqnarray}
For the states $n=1$, the coefficient $c_2$ has a simple expression
in terms of the BS amplitude:
\begin{equation} \label{c_2n=1}
c_2(n=1)=\frac{1}{4\pi^2}\frac{[g_{1\kappa}^{0\,\prime}(-1)]^2}{N_{tot}},
\end{equation}
where $g'(-1)$ is the derivative of $g$ with respect to $z$ at $z=-1$
and $N_{tot}$ is the normalization factor that ensures the condition
$F_e(0)=1$ when $g$ is arbitrarily normalized (its expression is given
in Eq. (\ref{Ntot_1})). The expression of $c_0$ is more complicated and
depends on the bound state mass $M$, as well as on the function
$g$ over the whole region of $z$ in the interval $[-1,+1]$.

For the normal state $n=1,\kappa=0$, $c_0$ is negative in general,
but changes sign and becomes positive for small binding energies.
The expression of $g$ takes a simple
form in the two extreme cases of non-relativistic limit and maximal
binding energy ($M=0$). Normalizing $g$ so that $g(0)=1$,
one has in the first case $g(z)=(1-|z|)$, with
$N_{tot}=1/(32\pi\alpha^5)$, and in the second case $g(z)=(1-z^2)$,
with $N_{tot}=1/(270\pi^2)$ \cite{wick,cutk}. One obtains, in these
two extreme cases, the asymptotic behaviors \cite{dshvk}
\begin{eqnarray}
\label{ff_NR}  
F_e(Q^2)&\simeq& \frac{8\alpha^5}{\pi}\Big({m^2\over Q^2}\Big)^2
\left[\ln \left({Q^2\over m^2}\right) + \frac{2\pi}{\alpha} \right]
\nonumber\\
& & (\mathrm{non-relativistic\ limit}\; M\to 2m), \\ 
\label{ff_M0}
F_e(Q^2)&\simeq& 270 \Big({m^2\over Q^2}\Big)^2
\left[\ln \left({Q^2\over m^2}\right)- 4\right] \nonumber \\ 
& & (\mathrm{ultra-relativistic\ case}\;  M=0).
\end{eqnarray}
Notice that in Eq. (\ref{ff_NR}), we have neglected
$\alpha$-indepen\-dent
constant factors in front of the additive term  $2\pi/\alpha$.
The fact that $c_0$ is generally negative, except for small binding
energies, where it can, however, take a large value (proportional to
$1/\alpha$), has as a main consequence the screening of the logarithmic
tail, requiring, for a numerical analysis of the asymptotic
behaviors, very large values of $Q^2$ ($Q^2\gg 100\div 1000\ m^2$). 

In order to put in evidence the asymtptotic behavior (\ref{Fe_asymp})
and to determine its leading terms we have computed the
``reduced form factor'' $\bar{F}_e(Q^2)$, defined as 
\begin{eqnarray}\label{Coefs_asymp}  
\bar{F}_e(Q^2) &\equiv& {1 \over \ln\left({Q^2\over m^2}\right)  }  \;
\Big(\frac{Q^2}{m^2}\Big)^2 \; F_e(Q^2) \nonumber\\
&_{\stackrel{{\displaystyle=}}
{Q^2\to\infty}}&
c_2 +  {c_0 \over \ln\left({Q^2\over m^2}\right)},  
\end{eqnarray}
which should  tend, when $Q^2\to \infty$, to a
constant (up to logarithmic corrections).
The asymptotic coefficients $c_i$ can be extracted from $\bar{F}_e(Q^2)$
and its derivative at a given $Q^2$ with the relations
\begin{equation}\label{c_0}
  c_0=  - \; { d\bar{F}(Q^2)\over dQ^2}  \; Q^2
  \ln^2\left({Q^2\over m^2}\right)
\end{equation}
and 
\begin{equation}\label{c_2}
c_2=   \bar{F}(Q^2) -  {c_0 \over \ln\left({Q^2\over m^2}\right)}  
\end{equation}

\begin{figure}[h!]
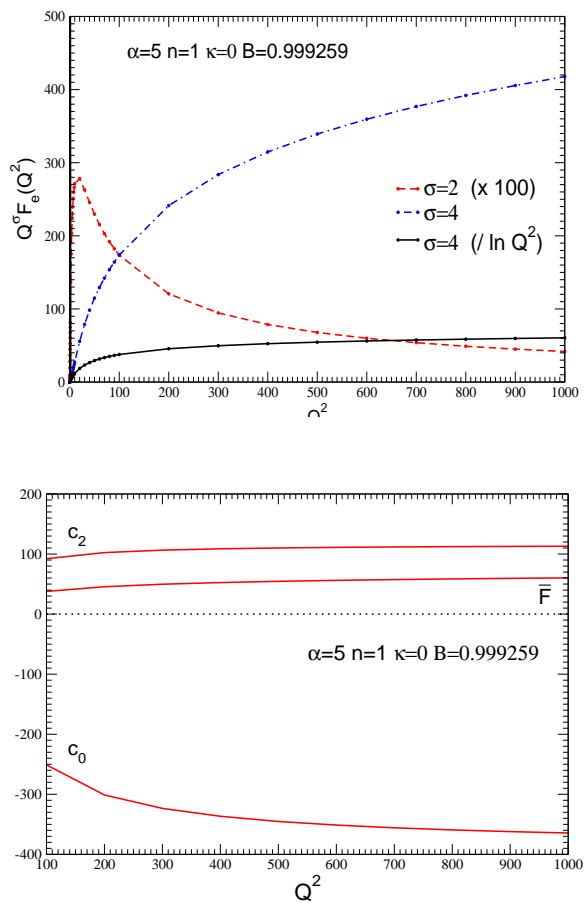

\begin{center}
\vspace{0.7cm}
\epsfxsize=7.5cm\epsfysize=5.5cm\epsfbox{QnuFQ2_Nb1_1000.eps}
\vspace{0.8cm}\\
\epsfxsize=7.5cm\epsfysize=5.5cm\epsfbox{c0_c2_Nb1_1000_NON.eps}
\end{center}
\caption{Asymptotic behaviours of the elastic form factor of the $n=1$
state No. 1 of Table \ref{tab1}. Upper panel: the elastic form factor
$F_e$ multiplied by $Q^{\sigma}$ with $\sigma=2$ (red), $\sigma=4$  (blue), and $\sigma=4$
divided by $\ln(Q^2)$.  
Lower panel: the  asymptotic coefficients  $c_0$ and $c_2$
defined in Eq. (\ref{Coefs_asymp}).}\label{QnuFQ2_1}
\end{figure}

The results for the $n=1$ state No. 1 from Table  \ref{tab1} are
displayed in Fig. \ref{QnuFQ2_1}.
The upper panel represents the elastic form factor multiplied by
$Q^{\sigma}$ with $\sigma=2$ (red line), $\sigma=4$ (blue line)
and  $\sigma=4$ divided by the logarithmic term, as in Eq.
(\ref{Coefs_asymp}), to exhibit the asymptotic behaviour derived in
Eq. (\ref{Fe_asymp})  
and the important contribution that the logarithmic term can have
in the  domain ${Q^2}\in[0,1000]$, even at $Q^2 \sim 1000$. The
latter is seen in the difference between the blue and black lines,
which exactly correspond to $\bar{F}_e$.
In the lower panel we have plotted the coefficients $c_0$ and $c_2$
in the ``asymptotic''  domain ${Q^2}\in[100,1000]$, together
with the full reduced form factor $\bar{F}_e$. They already show a nice
convergence at ${Q^2}=1000$, but the difference between $\bar{F}_e$
and its asymptotic value $c_2$ remains sizeable, due to the large
contribution of $c_0$, which decreases very slowly. We have also
checked the stability of our results with respect to the number
of grid points ($n_{grid}=400,800,1600$) used in computing the form factors:
the sensitivity is not visible by eyes and is not significant in our analysis.

\begin{figure}[h!]
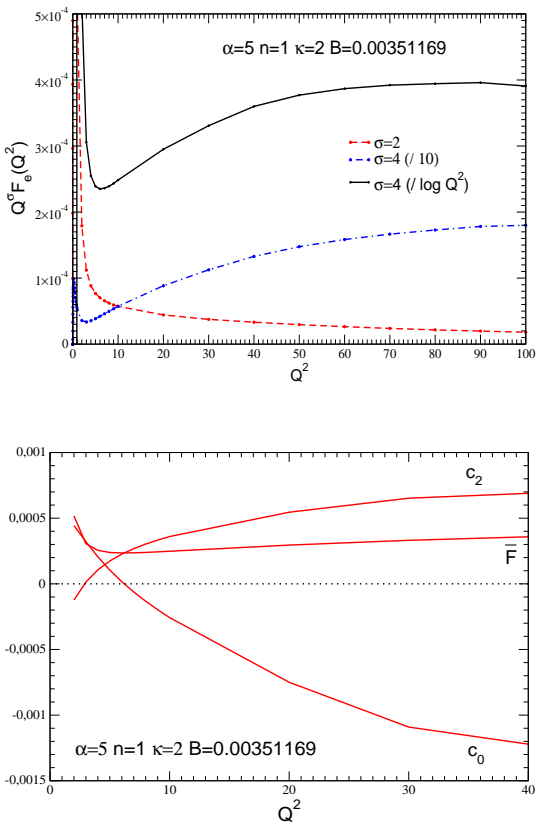

\begin{center}
\vspace{0.5cm}
\epsfxsize=7.cm\epsfysize=5.cm\epsfbox{QnuFQ2_Nb3.eps}\vspace{0.8cm}\\
\epsfxsize=7.cm\epsfysize=5.cm\epsfbox{c0_c2_Nb3_100.eps}
\end{center}
\caption{Same as in Fig. \ref{QnuFQ2_1}, but for the  $n=1$ abnormal
state No. 3 from Table \ref{tab1} .}\label{QnuFQ2_3}
\end{figure}

Fig. \ref{QnuFQ2_3} contains the same results for the $n=1$ abnormal
state No. 3.
Due to the faster decrease of the corresponding elastic form factor,
(see the lower panel of Fig. \ref{Fel_1_3}) the asymptotic regime is
reached at $Q^2\approx 50$, with the asymptotic constant
$c_2\approx 0.0004$, which is
seven orders of magnitude smaller than for the normal state No. 1.

Figures \ref{Fel_2} and \ref{Fel_4}  contain the elastic form factors
of the two $n=2$  states.
The result for the normal state  No. 2 ($\kappa=0, B=0.2084$) is
represented in the upper panel of  Fig. \ref{Fel_2}.
One first observes a much faster decrease than for the $n=1$ state No. 1
(upper panel of Fig. \ref{Fel_1_3}), having comparable binding energies:
one order of magnitude at $Q^2 =1$.
One also remarks the appearance of two zeroes in the form factor, which
becomes negatives in the range $Q^2\in[1.5,3.0]$. This is
a consequence of the complex structure of the state.
Indeed, the different  contributions $F^{\nu\nu'}$ depending on
$g_{20}^\nu g_{20}^{\nu'}$  are indicated in the lower panel.
As one can see, the physical $F_e$ results from strong cancellations
of terms which have opposite signs and are one order of magnitude larger
than the physical value they build. 
Although these components are of the same order, the contribution due
to $g^0_{20}$ -- which determines the binding energy  of the state --
is far from being dominant.

\begin{figure}[h!]
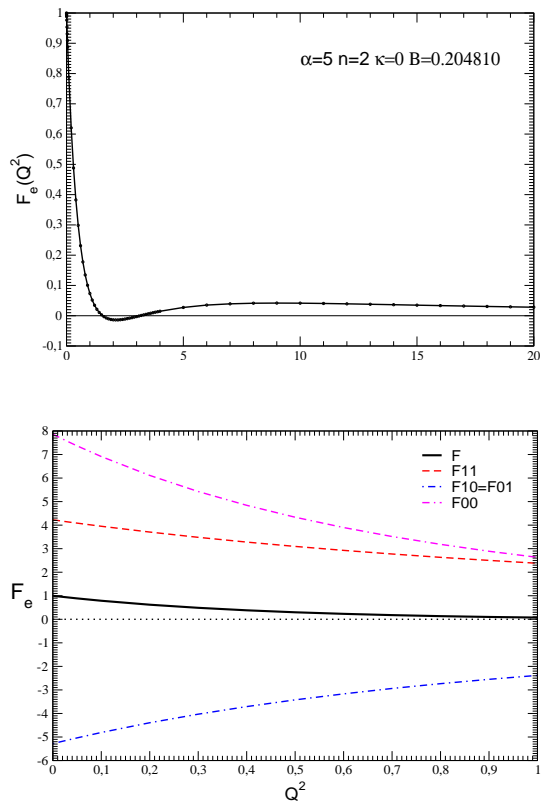

\vspace{0.6cm}
\begin{center}
\epsfxsize=7.cm\epsfysize=5cm\epsfbox{FQ2_Nb2_20.eps}\vspace{0.5cm}\\
\epsfxsize=7.cm\epsfysize=5cm\epsfbox{FQ2_Nb2_Dec.eps}
\caption{Upper panel: elastic form factor of the normal state No. 2 of
Table \ref{tab1}  ($n=2$, $\kappa=0$). The different  contributions
$F^{\nu\nu'}$ depending on $g_{20}^\nu g_{20}^{\nu'}$  are indicated in the lower
panel.}
\label{Fel_2}
\end{center}
\vspace{0.5cm}
\end{figure}

The elastic form factor of the abnormal state No. 4
($\kappa=2, B=0.00112$) is displayed in Fig. \ref{Fel_4}.
The same remark concerning the faster decrease than the n=1 state
No. 3 with comparable binding energy holds.
Notice also the non trivial structure -- similar to a diffraction
pattern --  seen at $Q^2\approx 0.01$ and detailed in the lower panel.
Such a structure, as well as the zeroes in upper panel of Fig \ref{Fel_2},
is totally unusual in a two-scalar system interacting  by the simple
kernel (\ref{ladder}) and indicates the complexity of the  wave function
for any state solution with $n>1$, be it normal or abnormal.
The decomposition of $F_e$ in terms of the different components
$F^{\nu\nu'}$ is similar than for the state No. 2, i.e., strong
cancellations occur among opposite sign larger terms.  
 
\begin{figure}[h!]
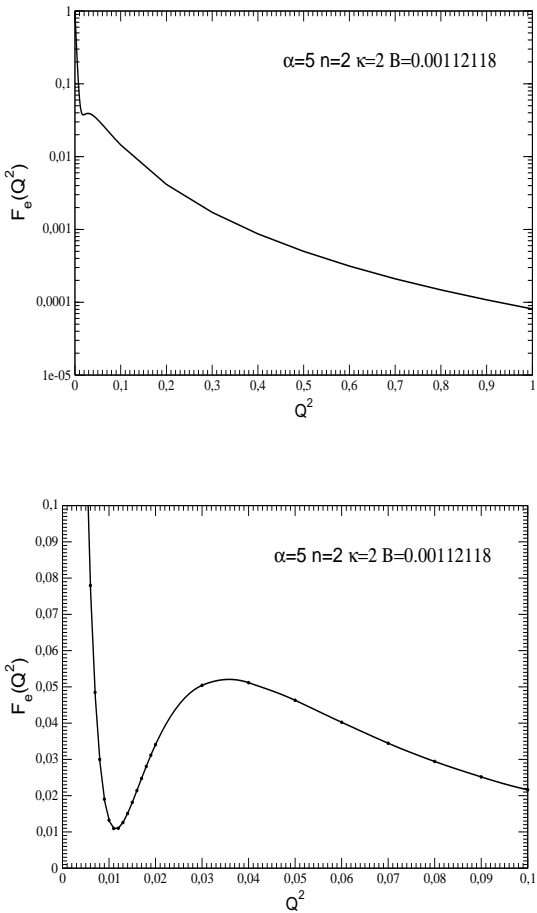

\vspace{0.6cm}
\begin{center}
\epsfxsize=7cm\epsfysize=5.5cm\epsfbox{FQ2_Nb4.eps}\vspace{1cm}\\
\epsfxsize=7cm\epsfysize=5.5cm\epsfbox{FQ2_Nb4_Zoom2.eps}
\caption{Elastic form factor of the abnormal state No. 4 of Table
\ref{tab1}  ($n=2$, $\kappa=2$). The non-trivial structure  at
$Q^2\approx 0.01$ is detailed in the lower panel.}\label{Fel_4}
\end{center}
\end{figure}

The corresponding rms radii, extracted using Eq. (\ref{r2}), are
$R_2=3.8$ fm (state No. 2) and $R_4=49.0$ fm (state No.4).
Just on the basis of their binding energies one should expect
twice smaller values.
The reason is again the complex structure of the BS amplitude  of
$n>1$ states, with a  $\nu>0$  dominating component (by a factor of
$10^3$ in state No. 4)  that plays no role in determining the
binding energy -- and so for the spatial extension --  of the system.

\begin{figure}[h!]
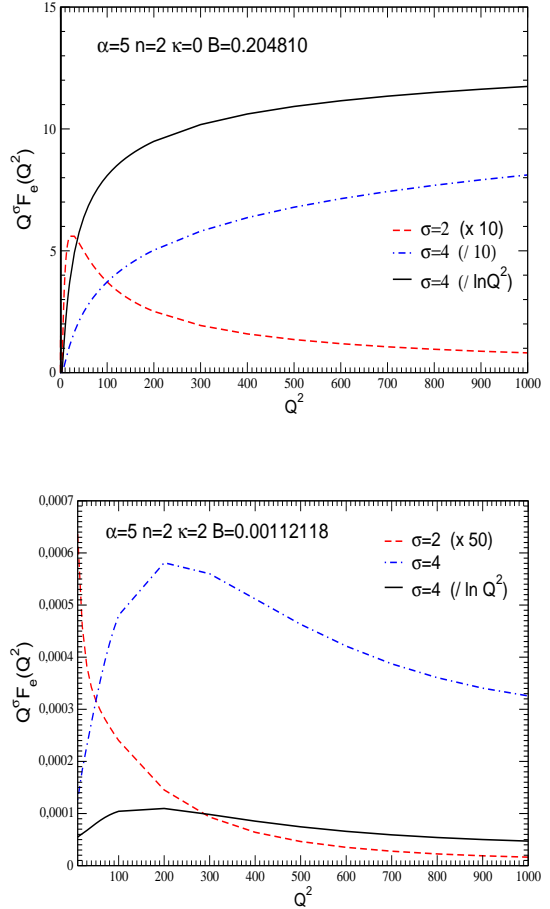

\vspace{0.6cm}
\begin{center}
\epsfxsize=7.cm\epsfysize=5.5cm\epsfbox{QnuFQ2_Nb2_1000.eps}\vspace{1cm}\\
\epsfxsize=7.cm\epsfysize=5.5cm\epsfbox{QnuFQ2_Nb4_1000.eps}
\end{center}
\caption{Asymptotic behaviour of the elastic form factors of $n=2$
states of Table \ref{tab1} (solid lines): No. 2 (normal) on top and No. 4 (abnormal) at bottom.}\label{QnuFQ2_24}
\end{figure}

As it was the case for the $n=1$ states, the comparison of the elastic
form factors of the $n=2$ normal and abnormal
states  (upper  panels of Figs. \ref{Fel_2} and \ref{Fel_4}) shows that
the abnormal state form factors decrease  faster than the normal ones
as functions of $Q^2$.
At $Q^2=1$ the ratio normal/abnormal is three orders of magnitude.
Even after adjusting the coupling constant of state No. 2, to have
the same binding energy than the state No. 4, the conclusion remains
unchanged.

It is also interesting to examine the asymptotic behaviour, which is
supposed to have the same form (\ref{Coefs_asymp}) than for $n=1$ states.
This is done in the two panels of Figure  \ref{QnuFQ2_24}. 
They show again the importance of logarithmic corrections and the
different orders of magnitudes of the asymptotic constant $c_2$ between
normal and abnormal sates. Notice the scaling factor introduced in some
of the plots to include the comparison in the same frame.

The ratios of form factors for the normal states Nos. 1 and 2 and for
the abnormal ones 3 and 4  are shown in Fig. \ref{rat1234} (upper panel).
They indeed tend to constants. The value of this constant is $\sim 7$.
The ratio of form factors for the abnormal states No. 3 and the normal
one No. 1 is shown in the lower panel of the same figure. It also tends
to a constant. The value of this constant is $\sim 10^{-5}$. 
Note that the ratio of form factors of different nature (abnormal/normal)
is much smaller than normal/normal and abnormal/abnormal, as expected.
Surprisingly, the ratios normal/normal and abnormal/abnormal are the same.
These asymptotic behaviors of the elastic form factors bring additional
arguments in favor of the interpretation of the abnormal states of the
W-C model as hybrids.   

\begin{figure}[h!]
\begin{center}
\epsfxsize=7.cm\epsfysize=5cm \epsfbox{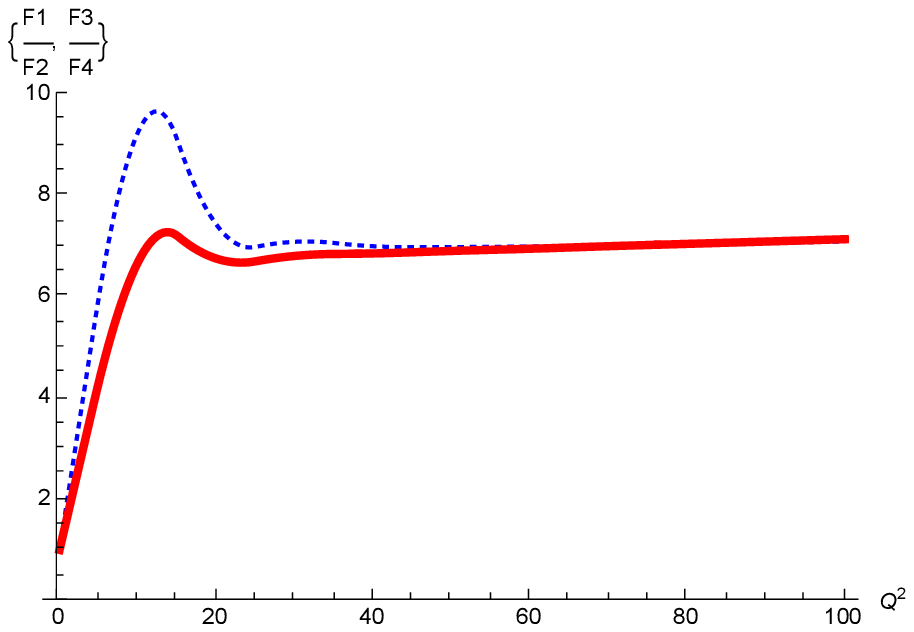}   \vspace{0.5cm}\\
\epsfxsize=7.cm\epsfysize=5cm\epsfbox{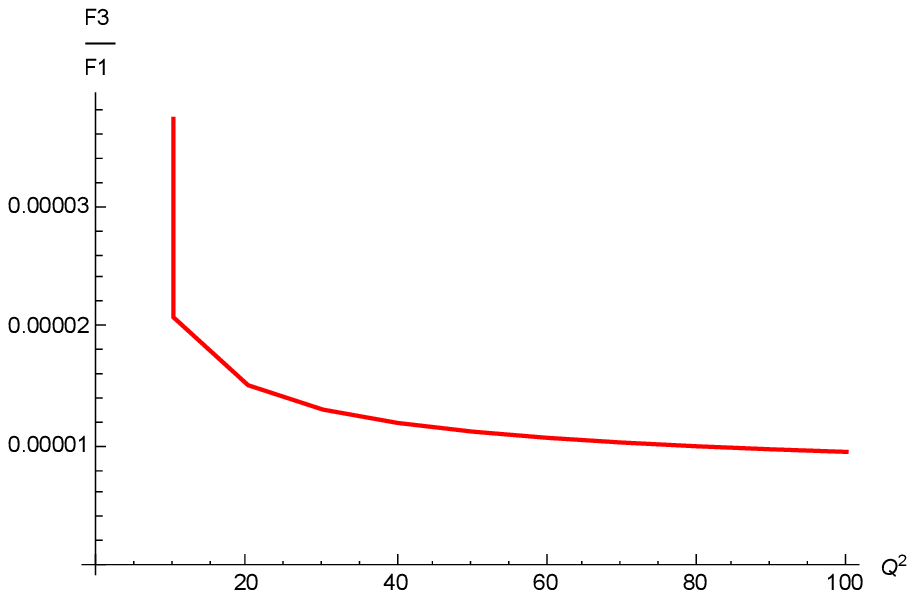}  
\caption{ (Color online) 
Upper panel: Dotted curve is the ratio of the form factors $F(Q^2)$ for
the normal ground state No. 1 and for the normal excited state No. 2. 
Solid curve is the same for the abnormal states No. 3 and No. 4.
Lower panel: The ratio of the form factors $F(Q^2)$ for the abnormal
state No. 3 from Table 1 and the normal one No. 1.}\label{rat1234}
\end{center}
\end{figure}
 
\subsection{Transition form factors $F_{if}(Q^2)$}\label{trF}

For the sake of completness in the study of abnormal solutions of the
W-C model we present here the results for the transition form factors.
There are four states in Table \ref{tab1} and, hence, six possible
transitions between them. 
The corresponding transition form factors $F$ are shown in Figs.
\ref{Ftr1_234} and \ref{F34_23_24}.
The comparison reveals a hierarchy of the transition form factors.
In Fig. \ref{Ftr1_234} we can see that the form factor for the
transition between two normal states, No. 1
($n=1,\kappa=0$) $\to$ No. 2 ($n=2,\kappa=0)$  (upper panel), dominates, 
by a factor $\sim 100$, over the maximal values of the normal $\to$
abnormal transitions (central and lower panels).

The transition form factor $F(Q^2)$ between two abnormal states, No. 3
($n=1,\kappa=2$) $\to$ No. 4 ($n=2,\kappa=2$), is displayed in Fig.
\ref{F34_23_24} (upper panel). Its maximal value has the same
order of magnitude as the normal-normal one, 
though it decreases much faster. At last, the form factors for the
transitions between the normal and abnormal states 
(Figs. \ref{Ftr1_234} and \ref{F34_23_24}, both central and  bottom
panels) are approximately 100 times smaller than the
abnormal$\leftrightarrow$abnormal form factor. This hierarchy is
apparently related to the need of rebuilding the state structure for
the normal$\leftrightarrow$abnormal transitions.

At  first glance, this dominance can be simply due to the very different
binding energies: two such states will have a small overlap,
without invocating any abnormal character. Indeed, one can hardly
separate unambiguously the effect of different structures from the
different binding energies (the latter result in different wave functions).
That is why we have carried out the complex analysis based on the behavior
of the form factor (elastic and inelastic) and on the content of the
Fock sectors.

\begin{figure}[h!]
\begin{center}
\epsfxsize=5.cm\epsfysize=4.cm\epsfbox{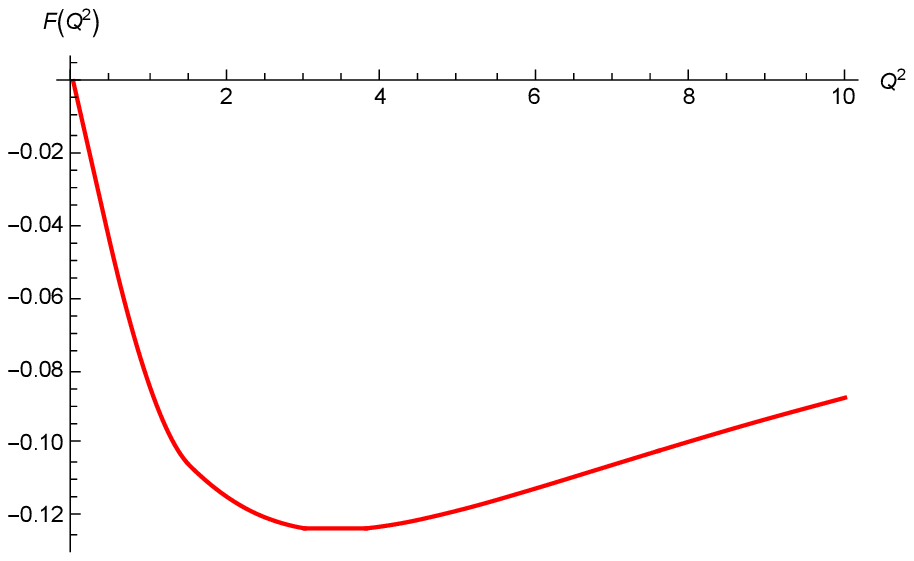}\vspace{0.5cm}\\
\epsfxsize=5.cm\epsfysize=4.cm\epsfbox{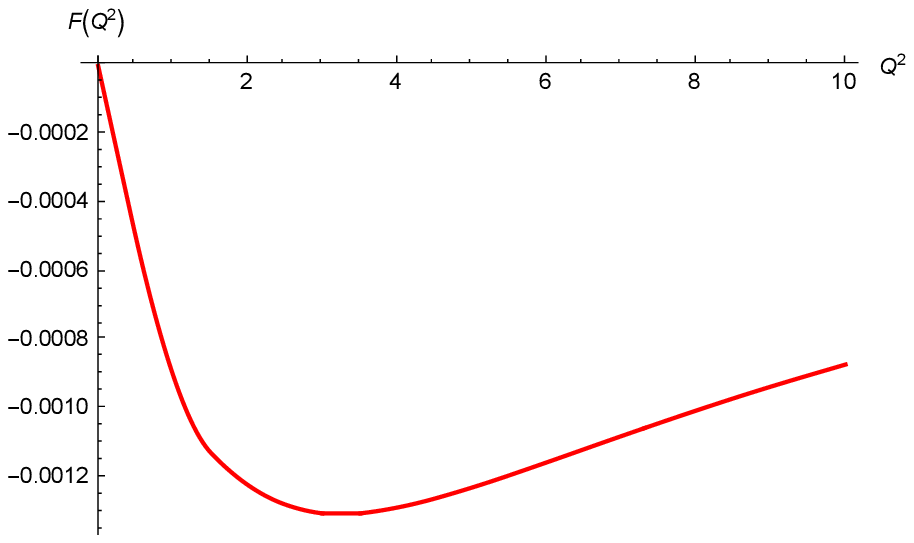}\vspace{0.5cm}\\
\epsfxsize=5.cm\epsfysize=4.cm\epsfbox{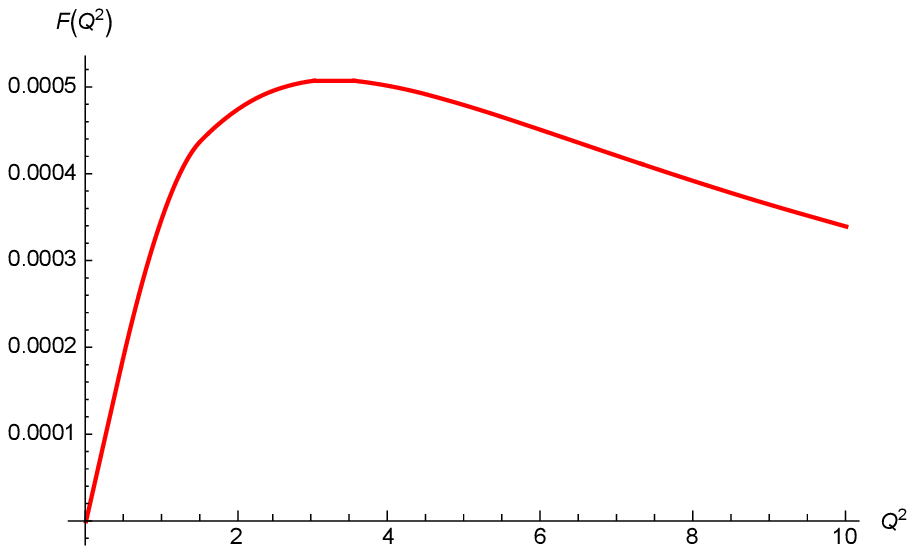}
\caption{Transition form factors  from  normal state  No. 1
($n=1,\kappa=0$) to other states listed in Table \ref{tab1}.
$1\to2$:     No.2 ($n=2, \kappa=0$, normal)  (upper panel);
$1\to3$:    No. 3 ($n=1, \kappa=2$,  abnormal)  (central panel) and
  $1\to4$:   No.4  ($n=2,\kappa=2$, abnormal) (lower panel).}
\label{Ftr1_234}
\end{center}
\end{figure}

\begin{figure}[h!]
\begin{center}
\epsfxsize=5.cm\epsfysize=4cm\epsfbox{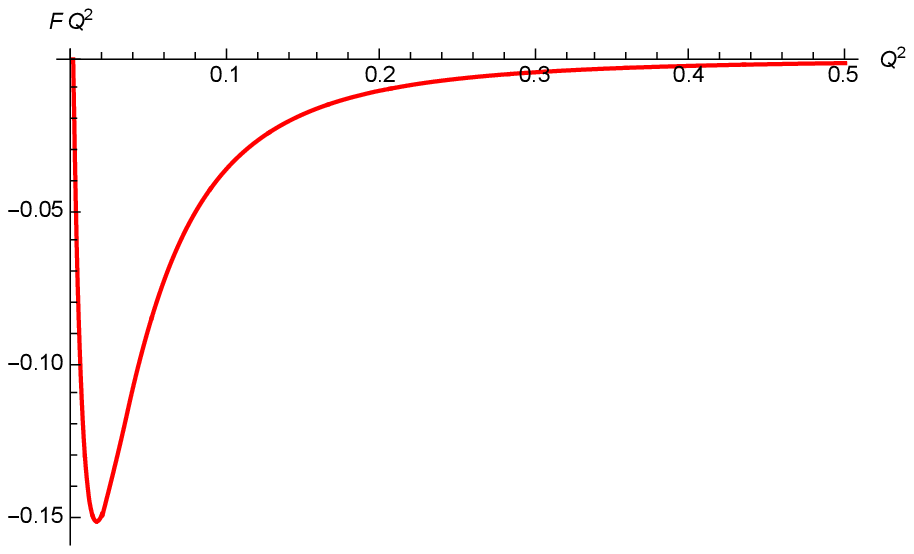} \vspace{0.5cm}\\
\epsfxsize=5.cm\epsfysize=4cm\epsfbox{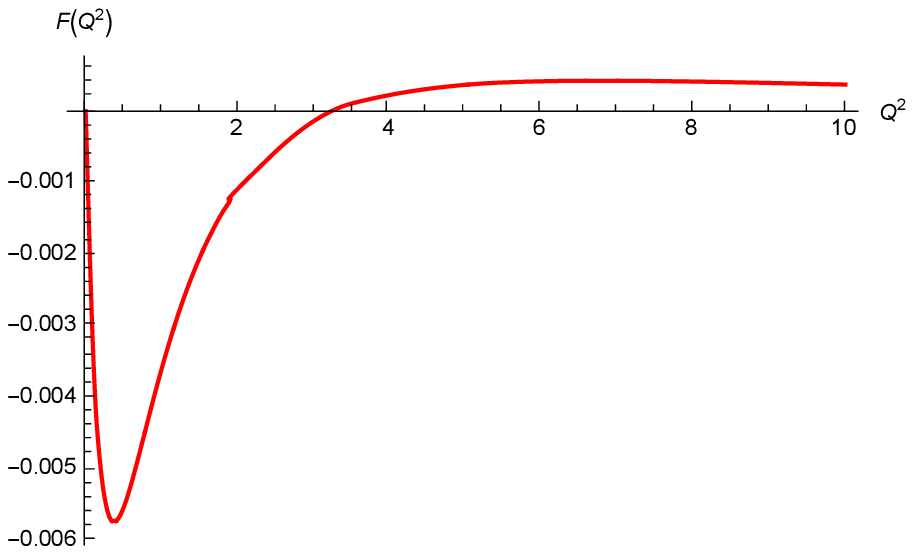}\vspace{0.5cm}\\
\epsfxsize=5.cm\epsfysize=4cm\epsfbox{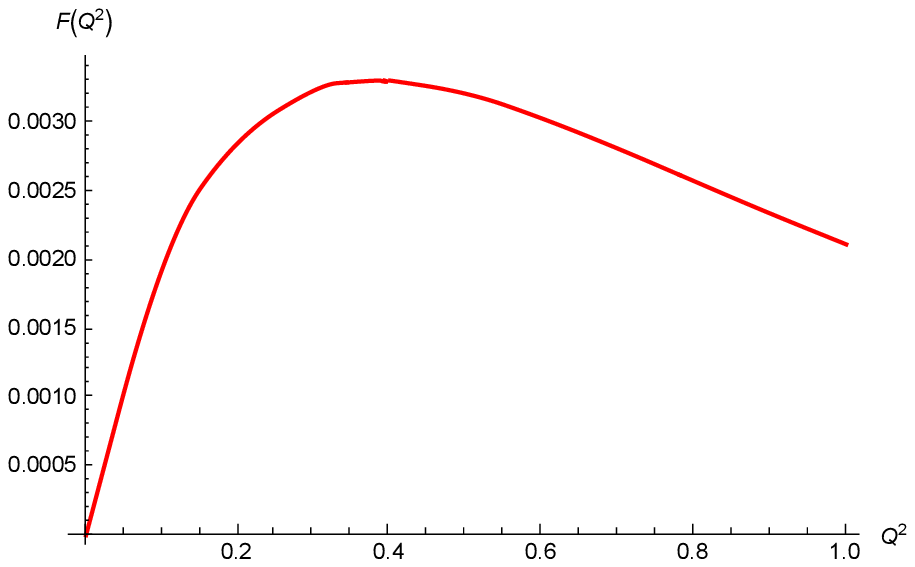}
\caption{Same as in Fig. \ref{Ftr1_234} between 
$3\to 4$:   No. 3($n=1,\kappa=2$, abnormal) and No. 4 ($n=2,\kappa=2$,
abnormal)
(upper panel); $2\to 3$: the  state No. 2 ($n=2,\kappa=0$, normal) and
the No. 3 ($n=1,\kappa=2$, abnormal)  (central panel) and
$2\to 4$: the  state  No. 2 ($n=2, \kappa=0$, normal) and  
No. 4 ($n=2,\kappa=2$, abnormal) (lower panel).}\label{F34_23_24}
\end{center}
\end{figure}

For all the transitions that were considered in this  section we have
calculated simultaneously  the transition form factor $G(Q^2)$.
The contraction of the electromagnetic current $J$ with the momentum
transfer $q$ results in $q\cd J=Q_c^2G(Q^2)$ [Eq. (\ref{FFp})].
Therefore, the current conservation implies $G(Q^2)\equiv 0$ for any $Q^2$
[Eq. (\ref{G})]. A formal proof of this equality, using the BS
equation, is given in \ref{proofG}.
Computing a quantity that we know from the first principles (current
conservation) that should be identically zero could be in principle
considered, at most,  as being superfluous.
However, as is seen from the derivation given in  \ref{proofG}, 
this property is directly related to the fact that $g$ is
indeed a solution of the BS equation
and the form factors,  mainly  the 3D integrals (\ref{Fe_kkp}) and
(\ref{Fif_kkp}), have been accurately computed.
It thus constitutes a test for our numerical solutions.
Similar tests were successfully carried out in Ref. \cite{tff} for the
form factors corresponding to the electro-desintegration of a bound
system (the transition discrete $\to$ continuous spectrum).

The kind of results obtained in computing $G(Q^2)$ is illustrated in Figs.
\ref{G_23}, in a single example corresponding to the transition between  
No. 2 $(n=2,\kappa=0)$ $\to$ No. 3 $(n=1,\kappa=2)$ states.
As in the elastic case, the form factor $G(Q^2)$ results from the sum
of two terms:  $G^{10}(Q^2)$  (proportional to $g_2^1g_1^0$ ) and
$G^{00}(Q^2)$ (proportional to $g_2^0g_1^0$). They are indicated
respectively at upper panel in dashed ($G^{10}(Q^2)$) and 
in dotted ($G^{00}(Q^2)$) lines. The sum of them, i.e., the full form
factor $G(Q^2)$, is indicated by a thick solid line and is
indistinguishable from zero at the scale of the figure.
It is in fact $\approx 10^{-6}$.

Note however that the sensitivity of $G(Q^2)$ to the accuracy in solving
the BS equation, that is, in computing the functions $g(z)$ and in
calculating the 3D integrals  (\ref{Fif_kkp}), is very high. 
The result $G(Q^2)\equiv 0$ is due to delicate cancellations
between terms which are several orders of magnitude greater.
A small error in these calculations results in non-zero $G(Q^2)$. This
is  demonstrated in the lower panel of Fig. \ref{G_23}.
Thus, an error in computing the binding energy, e.g.,  
setting $B_i= 3.512 \cdot 10^{-3}$ instead of $B_i= 3.51169 \cdot 10^{-3}$
and $B_f=0.2084$ instead of $B_f = 0.2084099$, provides
$G(Q^2) \approx -0.005$.
This value is comparable with the maximum value of the corresponding
form factor $F(Q^2)$ (in central panel of Fig. \ref{F34_23_24}).
An apparent violation of current conservation would hide in fact a lack
of accuracy in the computational procedure.

\begin{figure}[h!]
\begin{center}
\epsfxsize=6.5cm\epsfysize=5cm\epsfbox{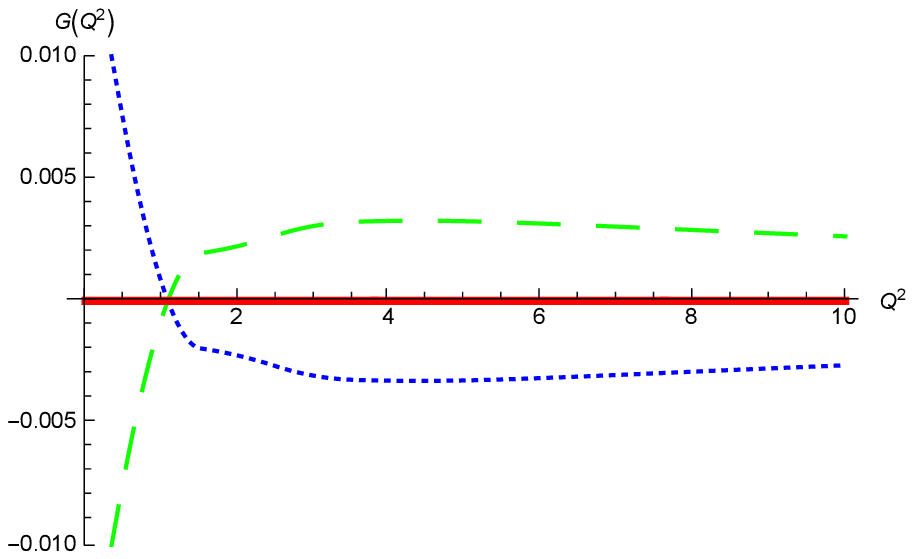}\vspace{0.5cm}\\
\epsfxsize=6.5cm\epsfysize=5cm\epsfbox{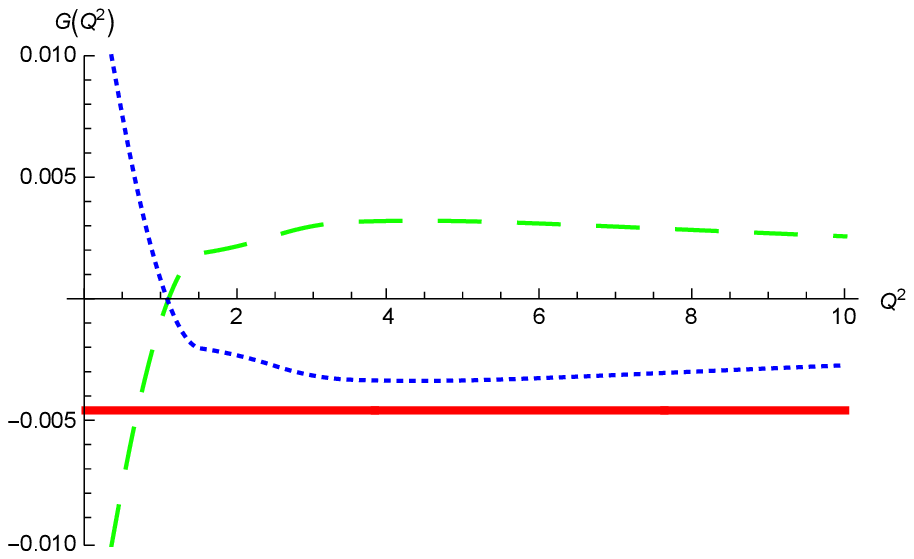}
\caption{(Color online) Contributions to the $2\to 3$ transition
form factor $G(Q^2)$: 
the dashed line is   $G^{10}(Q^2)$ and  the dotted line  $G^{00}(Q^2)$.
The sum of them  (solid line) is the full form factor $G(Q^2)$
(upper panel):
it is of the order of $10^{-6}$ and results from contributions of
several orders of magnitude greater.
A small error in computing the binding energy, e.g.,  
$B_i= 3.512 \cdot 10^{-3}$ instead of $B_i= 3.51169 \cdot 10^{-3}$
and $B_f=0.2084$ instead of $B_f = 0.2084099$, would provide
$G(Q^2) \approx -0.005$  (lower panel), a size comparable with  $F(Q^2)$ in
Fig. \ref{F34_23_24}.}\label{G_23}
\end{center}
\end{figure}

To summarize this section, the comparison of the elastic form factors
presented in Figs. \ref{Fel_1_3}, \ref{Fel_2}, \ref{Fel_4},
of the normal states  with the abnormal ones,
shows that the elastic form factors of the abnormal states vs.
$Q^2$ decrease much faster than for the  normal ones.
At $Q^2\sim 1$, the abnormal form factors are about $10^3$ times
smaller than the normal ones, whereas the behaviors of the elastic
form factors of the normal states with $n=1$ and $n=2$ remain very
close to each other. These observations confirm that the abnormal
states are dominated by the many-body Fock states
\cite{matvmurtavk,brodsfarr,radyush}. 
\par
The transitions between the normal and abnormal states, in comparison
to the normal-normal and abnor\-mal-abnormal transitions, are also
suppressed. This suppression indicates that the normal and abnormal
states have different structures and the transitions between them
require the rebuilding of the states.
\par
The quality of our numerical calculation is quite sufficient to
justify the above conclusions. This is demonstrated in  
Fig. \ref{G_23} for the transition form
factor $G(Q^2)$. As is explained in Sec. \ref{FFs} and proved in
 \ref{proofG}, the electromagnetic current conservation
requires $G(Q^2)=0$. This is indeed observed in Fig. \ref{G_23}, top,
(solid line) as a result of rather delicate
cancellations of several contributions. 
Numerical changes of $B$ and $g(z)$, which seem insignificant,
may noticeably change the value of $G(Q^2)$ (Fig. \ref{G_23}, bottom). 

The calculations of the form factors for the transitions to the states
given in the Table \ref{tab2}, with the precision used so far, are
unstable and require much higher precision. We do not present them here.

\section{Concluding remarks}\label{concl}

Our present analysis shows that the abnormal solutions
of the W-C model have a different internal structure than
the normal ones, which can be traced back to their decomposition
properties into Fock space sectors on light-front planes. 
This constitutes a genuine property of these states and we propose it as
an alternative  characteristic to the traditional explanation in terms of
temporal degrees of freedom excitations.
Whereas the normal solutions are dominated by the two-body Fock sector
made of the two massive
constituents, the abnormal ones are dominated by the Fock sectors made of
the two massive constituents and  several or many massless exchange
particles. This feature is also manifested through the fast decrease
of the electromagnetic form factors of the abnormal states, signalling
their many-body compositeness. Therefore, the abnormal states do
not appear as pathological solutions of the BS equation, but rather as
solutions having specifically a relativistic origin, through the
dominance, in their internal structure, of the massless exchange
particles. 

Another particular feature of the abnormal solutions is the
relatively large value of the coupling constant needed for their
existence ($\alpha>\pi/4$). While the stability condition of the
W-C model also requires that $\alpha$ be bounded by the
upper value $2\pi$, the corresponding window of permissible values
does not belong to the domain of perturbation theory and
the question of the validity of the ladder approximation can be raised.
This question has been examined in Ref. \cite{alkofer} in the light
of the incorporation into the model of the renormalization effects.
It turns out that the above
domain of values of the coupling constant is incompatible with
a consistent treatment of such effects. The renormalization constants 
violate the inequalities which follow from the positivity conditions
coming from the spectral functions of the renormalized fields.

The latter result brings us to questioning the effect of the
higher-order multiparticle exchange diagrams. This problem has been
dealt with in Ref. \cite{jalsazdj} in a model of QED, where
the two massive constituents are static and tied at fixed positions in
three-dimensional space.
The abnormal solutions corresponding essentially to excitations of
degrees of freedom described by
the relative time variable  (or equivalently, of the relative energy
variable), this model should rather preserve their possible existence.
It turns out that in this configuration, the two-particle Green's
function is exactly calculable: it does not display any abnormal
type of  bound state in its structure; only the normal ground-state
is present in the spectrum.  On the other hand, the BS equation, in
the ladder approximation, still continues exhibiting abnormal
solutions. A similar conclusion is also obtained
from a different approach, based on a three-dimensional reduction
of the BS equation with the inclusion of multiparticle exchange
diagrams \cite{bijteb}.

The above considerations 
support, as mentioned, another
interpretation of the abnormal solutions of the W-C model.
It is possible that  excitations of the degrees of freedom described 
by the relative time variable correspond, from the point of view of
the Fock decomposition, to  filling of the higher Fock sectors.
These two interpretations may not contradict each other, but rather
be compatible.

In the W-C model, the even-relative-ener\-gy abnormal
solutions appear as theo\-retically acceptable states, {and are a
constitutive part of the corresponding  S-matrix. The fact that
they are dominated in Fock space by the many-body massless exchange
particles may suggest that they are a kind of ``hybrid'' states.
They might represent the Abelian scalar analogs of the hybrids that
are searched for in QCD, which are coupled essentially to a pair of
quark-antiquark and one or several gluon fields. Here, however,
the non-Abelian property of the gauge group, as well as the existence
of gluon self-interactions, make the latter states better adapted
for experimental, as well as theoretical, investigations. On the
other hand, the possible relevance of the multiparticle exchange
diagrams in the kernel of the BS equation remains a key ingredient
for the ultimate conclusion as to whether the W-C model
may have any experimental impact. Note, however, that it is natural
to expect that since the multiparticle exchanges add extra exchange
particles in the intermediate states, 
they do not reduce but rather increase the higher Fock components. 

In any event, in spite of the fact that the W-C model
is an oversimplified model, it nevertheless contains the phenomenon
of the particle creation and gives an interesting example of natural
generation of hybrid states.
Therefore, the hybrid systems can naturally exist in more sophisticated
field theories and be detected in appropriate experiments.

The question of a possible existence of abnormal states in the case
of massive-particle exchanges is currently under investigation by
the present authors.

\par
\bigskip

\begin{acknowledgements}
V.A.K. is grateful to the theory group of the Institut de Physique
Nucl\'eaire d'Orsay (IPNO) for the kind hospitality and financial
support during his visits.
H.S. acknowledges support from the EU research and innovation program
Horizon 2020, under Grant Agreement No. 824093.
\end{acknowledgements}
\par

\appendix

\section{Two-body contribution {\boldmath $N_2$} to the full norm}
\label{appN2}

According to Eq. (\ref{eq2}), the full normalization is the sum of
contributions of all the Fock sectors.  
If the two-body wave function ($\psi_2$ in Eq. (\ref{p})) is known,
then the contribution of the two-body sector in the full norm reads:
\begin{equation}\label{N2}
N_2=\frac{1}{(2\pi)^3}\int|\psi|^2(\vec{k}_{\perp},x)
\frac{d^2k_{\perp}dx} {2x(1-x)}.
\end{equation}
We recall that we follow the definition of the wave function of light-front dynamics,
where the state vector $|p\rangle$ is defined on a light-front plane.
More precisely, we use the explicitly covariant version of
light-front dynamics \cite{cdkm}, in which the light-front plane is
defined by the equation $\omega\cd x=0$, where $\omega$ is a four-vector
with the property $\omega^2=0$. 

The light-front wave function, which is the coefficient of 
the  two-body contribution $|2\rangle$ in the Fock decomposition
(\ref{p}), can be extracted from the BS amplitude
by projecting it on the light-front plane (see Eq. (3.57)
of Ref. \cite{cdkm}):
\begin{eqnarray}\label{lfwf}
\psi(\vec{k}_{\perp},x)=\frac{x(1-x)}{\pi\sqrt{N_{tot}}}
\int_{-\infty}^{\infty}\Phi\left(k+\frac{\beta\omega}
{\omega\cd p},p\right) d\beta,
\end{eqnarray}
where $N_{tot}$ is the full normalization factor (the same as in
Eqs. ({\ref{Fe_kkp}) and  ({\ref{Fif_kkp}) below) providing the
normalization of the elastic form factor $F(0)=1$.

Substituting here the BS amplitude (\ref{Phi2}) and making the
transformations resulting in Eq. (3.67) of Ref. \cite{cdkm}, we 
obtain, for $n=2$, an expression similar\footnote{Eq.
(\ref{wf1}) differs from Eq. (3.67) of Ref. \cite{cdkm} by a factor
$\sqrt{4\pi}$ because of different definitions of $g(z)$.} to
Eq. (3.67):
\begin{eqnarray}\label{wf1}
& &\psi(\vec{k}_{\perp},x)=-\frac{ix(1-x)}{\pi\sqrt{N_{tot}}}
\int_{-\infty}^{\infty}d\beta\int_{-1}^1 dz \nonumber \\
& &\ \ \ \ \ \ \times\frac{m^3g_2^1(z)}
{[(\vec{k}\,^2+\kappa_0^2)(1-zz_0)+\beta(z_0-z)-\imath \epsilon)]^3}
\nonumber\\
& &\ \ \ \ \ \ \ \ \
-\frac{ix(1-x)}{\pi\sqrt{N_{tot}}}\int_{-\infty}^{\infty}
d\beta\int_{-1}^1 dz
\nonumber\\
& &\ \ \ \ \ \ \times\frac{m^5g_2^0(z)}{[(\vec{k}\,^2+\kappa_0^2)(1-zz_0)
+\beta(z_0-z)-\imath\epsilon)]^4},
\end{eqnarray}
where 
\begin{eqnarray}\label{z0}
z_0&=&1-2x,\quad \kappa_0^2=m^2-\frac{1}{4}M^2, 
\nonumber\\
 \vec{k}\,^2
&=&\frac{\vec{k}\,^2_{\perp}+m^2}{4x(1-x)}-m^2.
\end{eqnarray}
To calculate these integrals, we use the formulas:
\begin{eqnarray}\label{delta}
&&\int_{-\infty}^{\infty}\frac{d\beta}{(\beta u+v-\imath \epsilon)^3}
=\frac{i\pi}{v^2}\delta(u),
\nonumber\\
&&\int_{-\infty}^{\infty}\frac{d\beta}
{(\beta u+v-\imath \epsilon)^4}=\frac{2i\pi}{3v^3}\delta(u).
\end{eqnarray}
In our case, 
$v=(\vec{k}\,^2+\kappa_0^2)(1-zz_0),\quad u=z-z_0$.
In this way, we find
\begin{eqnarray}\label{wf2}
\psi(\vec{k}_{\perp},x)
&=&\frac{m^3x(1-x)g_2^1(1-2x)}
{\sqrt{N_{tot}}[\vec{k}_{\perp}^2+m^2-x(1-x)M^2]^2}
\nonumber\\
&+&\frac{2m^5x(1-x)g_2^0(1-2x)}{3\sqrt{N_{tot}}
[\vec{k}_{\perp}^2+m^2-x(1-x)M^2]^3}.
\end{eqnarray}
Substituting $\psi(\vec{k}_{\perp},x)$
into the two-body normalization integral (\ref{N2}) and integrating,
one finds for $n=2$
\begin{eqnarray}\label{norm2}
N_2&=&\frac{1}{3\cdot 2^7\pi^2  N_{tot}}\int_{-1}^1dz\,(1-z^2)
\left\{\frac{[g_2^1(z)]^2}{[Q(z)]^3}\right.
\nonumber\\
&+&\left.\frac{g_2^1(z)g_2^0(z)}{[Q(z)]^4}
+\frac{4}{15}\frac{[g_2^0(z)]^2}{[Q(z)]^5}\right\},
\end{eqnarray}
with $Q(z)$ defined in Eq. (\ref{Q}).

\par
The case $n=1$, is obtained by  keeping  in the wave function 
(\ref{wf2}) the first term only and replacing in it $g_2^1$
by $g_1^0$. This gives:
\begin{equation}\label{norm10}
N_2=\frac{1}{384 \pi^2 N_{tot}}\int_{-1}^1\frac{(1-z^2)[g_1^0(z)]^2dz}
{[Q(z)]^3}.
\end{equation}

It is interesting to investigate analytically the behavior of $N_2$ in
the non-relativistic limit when the binding energy $B=2m-M$ tends to zero.
One can show that in this limit at leading order
$N_2(B\to 0)=1$ and then find the law of approaching of $N_2(B)-1$ to 0.
We will demonstrate this in the case $n=1$ [Eq. (\ref{norm10})].
$Q(z)$ [Eq. (\ref{Q})] can be represented as
$Q(z)\approx z^2+\frac{B}{m}$. Then
$$
\frac{1}{[Q(z)]^3}\approx \frac{1}{\left(z^2+\varepsilon\right)^3}\approx
\frac{3\pi}{8\varepsilon^{5/2}}\delta(z),\quad \varepsilon=\frac{B}{m}.
$$
We substitute this formula in (\ref{norm10}) and find the leading term
when $B\to 0$:
\begin{equation}\label{norm0}
N_2(B\to 0)=\frac{[g_1^0(0)]^2m^{5/2}}{2^{10}\pi N_{tot}B^{5/2}}.
\end{equation}

We then calculate $N_{tot}$ when $B\to 0$. 
For $n=1$, the sum (\ref{ff}) is reduced to one term only, determined
by  Eq. (\ref{Fe_1}) of  \ref{exprssff}.
$N_{tot}$ is given by Eq. (\ref{Ntot_1}).
Replacing in (\ref{Ntot_1}) the variable $\xi_+=\frac{1}{2}(1+y)$ and
keeping the leading term only, we transform the integrand as
\begin{eqnarray*}
\frac{m^2 (5-6\xi_+)   -2M^2\xi_+(1-\xi_+)}{[m^2-\xi_+(1-\xi_+)M^2]^4}
&\approx &\frac{2}{m^6(y^2+\varepsilon)^3}
\\
&\approx&
\frac{3\pi}{4\varepsilon^{5/2}}\delta(y).
\end{eqnarray*}
We omitted the derivative of the delta-function $\delta'(y)$ which
enters with a coefficient of the same order as $\delta(y)$, however,
it does not contribute for symmetric $g(z)$. Therefore, we get
\begin{eqnarray}\label{Ntot1}
N_{tot}&=&\frac{m^{5/2}}{2^7\pi B^{5/2}}\int_{-1}^1g(z)dz
\int_{-1}^1g(z')dz'\int_0^1du
\nonumber\\
&\times&u^2(1-u)^2 \delta\Bigl((1+z)u+(1+z')(1-u)-1\Bigr).
\nonumber\\
&&
\end{eqnarray}
From the delta-function we find $u=-z'/(z-z')$. This expression is in
the limits 
$0 \leq u \leq 1$, if $-1\leq z\leq 0$,  $0\leq z'\leq 1$ and
$0\leq z\leq 1$, $-1\leq z'\leq 0$. For symmetric $g(z)$ these two
domains give equal contributions. Therefore, integrating over $u$ by
means of the delta-function, we find for symmetric $g(z)$:
\begin{eqnarray}\label{norm00}
N_{tot}(B\to 0)&=&\frac{3m^{5/2}}{2^6\pi B^{5/2} }
\nonumber\\
&\times&\int_{0}^1\int_{0}^1
\frac{z^2 {z'}^2 g(z) g(z')}{(z+z')^5} dz dz'.
\end{eqnarray}
When $B\to 0$, the limiting expression of $g(z)$ is \cite{wick,cutk}:
$g(z)\approx 1-|z|$. Calculating the integral in (\ref{norm00}):
$$
\int_{0}^1\int_{0}^1 \frac{z^2 {z'}^2g(z) g(z')}{(z+z')^5} dz dz'=
\frac{1}{48},
$$
we get
$$
N_{tot}(B\to 0)=\frac{m^{5/2}}{2^{10}\pi\, B^{5/2}}.
$$
Substituting it in (\ref{norm0}), we obtain $N_2(B\to 0)=1$, as
should be for the normal state in the non-relativistic case.

The next order correction was found in Ref. \cite{dshvk}, Eq. (24).
It reads:
\begin{equation}\label{N2_1}
N_2-1=\frac{2\alpha}{\pi}\ln(\alpha).
\end{equation}
In the given order, we can use the Balmer formula:
$B=\frac{1}{4}m\alpha^2$, that is $\alpha=\sqrt{\frac{4B}{m}}$.
Substituting this $\alpha$ in (\ref{N2_1}), we find for $B\ll m$:
\begin{equation}\label{N2_1a}
N_2-1=\frac{1}{\pi}\sqrt{\frac{4B}{m}}\ln \left(\frac{4B}{m}\right).
\end{equation}

\section{Proof that $G(Q^2)\equiv 0$}\label{proofG}

The vanishing of the form factor $G$ for the transition between
different states is realized if the initial and final states are
associated with the solutions of the BS equation. Let us rewrite
the BS equation (\ref{bs}) so that it would contain the functions
$\Phi_i(\frac{1}{2}p - k,p)$ and $\Phi_f(\frac{1}{2}p' - k,p')$ entering
in the expression (\ref{ffbs}) for the current. The corresponding
equations, obtained from Eq. (\ref{bs}) by the variable shift 
$k\to \frac{1}{2}p-k$, are (assuming time-reversal and CP invariances)
\begin{eqnarray}\label{eqbs1}
&&\left[(p - k)^2 - m^2\right](k^2 - m^2)\Phi_i
\left(\frac{1}{2}p - k,p\right)
\nonumber\\
&&=16i\alpha\pi m^2 \int \frac{d^4k'}
{(2\pi)^4}\frac{\Phi_i\left(\frac{1}{2}p - k',p\right)}
{(k-k')^2+\imath\epsilon} \equiv K*\Phi_i,
\nonumber\\
&&
\\
&&\overline{\Phi}_f\left(\frac{1}{2}p' - k,p'\right)
\left[(p' - k)^2 - m^2\right](k^2 - m^2)
\nonumber\\
&&=16i\alpha\pi m^2 \int \frac{d^4k'}{(2\pi)^4}
\frac{\overline{\Phi}_f\left(\frac{1}{2}p' - k',p'\right)}
{(k-k')^2+\imath\epsilon} \equiv \overline\Phi_f*K,
\nonumber\\
&&
\label{eqbs2}
\end{eqnarray}
where the argument $\frac{1}{2}p - k$ is the relative momentum of the
particles having the momenta $p-k$, $k$ and similarly for
$\frac{1}{2}p' - k$.
Also notice that the kernel $K$ is the same in both equations.
\par
According to Eq. (\ref{FFp}):
$$
G(Q^2)=q\cd J/Q_c^2,
$$
with $q=p'-p$ and $J$ determined by Eq. (\ref{ffbs}).
The integrand of $q\cd J$ [cf. Eq. (\ref{ffbs})] contains the scalar
product 
$$
q\cd(p+p'-2k)= \Bigl({p'}^2-p^2-2(p'-p)\cd k\Bigr).
$$
On the other hand:
\begin{eqnarray*}
&&\Bigl({p'}^2-p^2-2(p'-p)\cd k\Bigr)(k^2-m^2)
\\
&&
\\
&=&\Bigl((p'-k)^2-m^2\Bigr)(k^2-m^2)
\\
&-&\Bigl((p-k)^2-m^2\Bigr)(k^2-m^2)
\\
&&
\\
&=&\left[\Bigl((p'-k)^2-m^2\Bigr)(k^2-m^2)-*K\right]
\\
&-&\left[\Bigl((p-k)^2-m^2\Bigr)(k^2-m^2)-K*\right].
\end{eqnarray*}
After substitution in $q\cd J$, the first operator acts on
$\overline{\Phi}_f$ and gives zero, and similarly with the second
operator acting on $\Phi_i$.
Hence 
$$
q\cd J=0 \qquad \Rightarrow \qquad G(Q^2)=0.
$$
\par
We emphasize that this result is obtained provided $\Phi_{i,f}$
satisfy the BS equation with a given kernel. Numerical uncertainties
in $\Phi_{i,f}$ result in deviations of $G(Q^2)$ from zero. This is
observed in Fig. \ref{G_23} (bottom).

\section{Expressions of the form factors}\label{exprssff}

In general,  the elastic and transition form factors are obtained as
a sum of $n\times n'$ terms $F^{\nu\nu'}$, involving the
different components of the BS amplitude in the initial ($n$)  and final
($n'$) states. They have the general form given in Eq. (\ref{ff}):
\begin{equation}\label{ffp} 
F(Q^2)= \sum_{\nu=0}^{n-1}\sum_{\nu'=0}^{n'-1} F^{\nu\nu'}(Q^2). 
\end{equation}
We present here explicit expressions for calculating the elastic
form factors in the cases $n=1,2$  and the transition form factors
between $n=2$ and $n'=2$ states.

The elastic form factor, $F_e$, for  $n=1$ states, determined by a single
component $g\equiv g_1^0$, is given by
\begin{eqnarray}\label{Fe_1}
F_e(Q^2)&=& {m^6\over 32\pi^2N_{tot}} \;  \int_{-1}^1 dz  \; g(z) \;
\int_{-1}^1 dz' g(z') 
\nonumber\\
&\times&   \int_0^1 du\; u^2(1-u)^2 \; {N \over D^4},    
\end{eqnarray}
with the integrands given by
\begin{eqnarray*}
N(z,z',u,Q^2) &=& A(z,z',u)+B(z,z',u)\; Q^2,   \cr 
 D(z,z',u,Q^2)     &=&  m^2 - M^2\xi_+(1-\xi_+)   -B(z,z',u)\; Q^2,   
\end{eqnarray*}
in terms of functions $A$ and $B$:
\begin{eqnarray*}
\xi_{\pm}(z,z',u)  &=&  {1\over2} \left[ (1+z)u \pm (1+z')(1-u) \right],
  \cr
A(z,z',u)   &=&   m^2(5-6\xi_+) -2M^2\xi_+(1-\xi_+),  \cr
B(z,z',u)   &=&  -{1\over4} (1+z)(1+z') u(1-u)\nonumber \\  
&=&-{1\over4} (\xi_+^2-\xi_-^2).
\end{eqnarray*}
The total norm $N_{tot}$, defined to fulfill $F(0)=1$, is obtained by 
\begin{eqnarray}\label{Ntot_1}
N_{tot}&=& {m^6\over 32\pi^2}  \  \int_{-1}^1 dz  \ g(z) \
\int_{-1}^1 dz' g(z')   \int_0^1 du\; u^2(1-u)^2 
\nonumber\\
&\times&\frac{m^2 (5-6\xi_+)   -2M^2\xi_+(1-\xi_+)}
{[m^2-\xi_+(1-\xi_+)M^2]^4}.
\end{eqnarray}

\bigskip
For $n=2$ states, the BS amplitude is determined by two components
$g_2^0,g^1_2$ (index $\kappa$ is here irrelevant and omitted).
The expressions of the different form factors $F_e^{\nu\nu'}$ can be written
in close analogy with (\ref{Fe_1}) and read
\begin{small}
\begin{eqnarray}\label{Fe_kkp}
F_e^{\nu\nu'}(Q^2) &=&\frac{m^6} {32\pi^2N_{tot} }  \int_{-1}^{+1} dz  \;
g_2^\nu(z)
\int_{-1}^{+1} dz' \; g_2^{\nu'}(z')   
\nonumber\\
&\times&\int_0^1 du\; u^{3-\nu} (1-u)^{3-\nu'}
\left({m^2\over D}\right)^{2-\nu-\nu'} {N^{\nu\nu'}  \over D^4 }, 
\nonumber\\
&&    
\end{eqnarray}
\end{small}
with the integrands expressed in terms of functions $A_{\nu\nu'}$ and
$B_{\nu\nu'}$ as 
\begin{equation}\label{Nij}
N^{\nu\nu'} =  A_{\nu\nu'}   \;+\; B_{\nu\nu'}\; Q^2,   
\end{equation}
with $A_{11}\equiv A$, $B_{11}\equiv B$ and the remaining ones given by
\begin{eqnarray}
A_{10}&=&    {1\over4}   
\left[ 4m^2 (6-7\xi_+) -4M_i^2\xi_+(2-\xi_+-\xi_+^2) \right], \cr
A_{00}&=&    {1\over3} 
\left[ 4m^2 (7-8\xi_+) -8M_i^2\xi_+(1-\xi_+^2)   \right],  \cr
B_{10}&=& -  {1\over4}  \left[ (2-\xi_+)
(\xi_+^2-\xi_-^2) \right], \cr
B_{00}&=& -  {1\over3}  \left[ (3-2\xi_+)
(\xi_+^2-\xi_-^2)   \right], 
\end{eqnarray}
and the symmetry relation $F^{10}=F^{01}$.
The total norm is given by
\begin{small}
\begin{eqnarray}
&&N_{tot}=  {m^6\over 32\pi^2}  \sum_{\nu,\nu'=0}^1
\int_{-1}^{+1} dz  \; g_2^\nu(z) \; \int_{-1}^{+1} dz' \; g_2^{\nu'}(z')
\nonumber\\
&\times&\int_0^1 du\; u^{3-\nu} (1-u)^{3-\nu'} \left.
\left({m^2\over D}\right)^{2-\nu-\nu'}
{N^{\nu\nu'}  \over D^{4} }\right|_{Q^2=0}.      
\nonumber\\
&& 
\end{eqnarray}
\end{small}
Notice that since $A_{11}\equiv A$, $B_{11}\equiv B$, $F_e^{11}(Q^2) $
from  Eq.  (\ref{Fe_kkp})   provides the case n=1.

The transition form factor between $n=1$ and $n=2$ states offers several
possibilities: $1\to 1$, $1\to 2$, $2\to 1$ and $2\to2$. 
We give below the expressions of the components $F_{if}^{\nu\nu'}$
of the transition form factors
between an initial state  $n=2$ and a final state $n=2$, determined,
respectively, by the components ($g_2^{0,i},g_2^{1,i}$) 
($g_2^{0,f},g_2^{1,f}$) and the total norms of the corresponding elastic
form factors $N_{tot}^i$, $N_{tot}^f$:
\begin{eqnarray}\label{Fif_kkp}
F^{\nu\nu'}_{if}&=&
{m^6\over32\pi^2\sqrt{N_{tot}^{i} N_{tot}^{f} } } \int_{-1}^{+1} dz
g_2^{\nu,i}(z) \int_{-1}^{+1} dz' g_2^{\nu',f}(z')
\nonumber\\
&\times&\int_0^1 du\;u^{3-\nu} (1-u)^{3-\nu'}
\left({m^2\over D}\right)^{2-\nu-\nu'}{N^{\nu\nu'}_{if}  \over D_{if}^{4}}.     
\nonumber\\
&&
\end{eqnarray}
The integrand $N^{\nu\nu'}_{if}$  and $D$ can be written in the form
\begin{eqnarray}
N^{\nu\nu'}_{if}(z,z',u)&=&  A^{\nu\nu'} + B^{\nu\nu'} \; Q^2 + C^{\nu\nu'} \; Q_c^2,
\label{N_if} \\
D_{if}(z,z',u)    &=& m^2-\xi_+ (1-\xi_+)M_i^2 
\nonumber\\
&&\phantom{ m^2-}- B^{11} \; Q^2  +
{1\over2} C^{11} \;Q_c^2, \label{Den_if}
\end{eqnarray}
involving now $Q_c^2={M_f}^2-M_i^2$ and new functions $C^{\nu\nu'}$.
The functions A and B are the same as for the elastic case with
$M^2\equiv M_i^2$ and the functions $C^{\nu\nu'}$ are defined as
\begin{eqnarray}\label{C}
C^{11}&=&  - (\xi_+-\xi_-)(1-\xi_+), \cr
C^{10}&=&  -2 {1\over4} 
(\xi_+-\xi_-)(2-\xi_+-\xi_+^2),  \cr
C^{00}&=&  - 4 {1\over3} 
(\xi_+-\xi_-)(1-\xi_+^2).  
\end{eqnarray}

The transition form factor $G(Q^2)$ is obtained by an expansion similar
to (\ref{ffp}). 
The corresponding components $G^{\nu\nu'}_{if}$ are given by
(\ref{Fif_kkp}), where the integrands $N_{if}^{\nu\nu'}$ are replaced by 
\begin{eqnarray}
\bar{N}_{if}^{\nu\nu'}(z,z',u)&=&  A^{\nu\nu'} \;+\; \bar{B}^{\nu\nu'}\; Q^2  \;+\;
C^{\nu\nu'}\; Q_c^2   
\nonumber\\
&&\phantom{A^{\nu\nu'} \;+\; \bar{B}^{\nu\nu'}\;}+{D}^{\nu\nu'}  \;
\left({Q^2\over Q_c^2}\right).
\label{barN_if}
\end{eqnarray}
Their form is similar to that of (\ref{N_if}), with the same $A^{\nu\nu'} $
and $C^{\nu\nu'} $,  but with two new coefficients $\bar{B}^{\nu\nu'} $
(instead of $B^{\nu\nu'} $) and $D^{\nu\nu'}$ given by:
\begin{eqnarray}
\bar{B}^{11}&=& - {1\over4}(5\xi_-^2-6\xi_+\xi_-+\xi_-^2),     \cr
\bar{B}^{10}&=&   {1\over4}   (\xi_+ - \xi_-)
[\xi_-(4+3\xi_+)- \xi_+(2-\xi_+)],   \cr
\bar{B}^{00}&=&  {1\over3}  (\xi_+ - \xi_-)
[ \xi_+(2\xi_+-3) + \xi_-(3+6\xi_+) ],    \cr
{D}^{11}&=&  + 3\xi_-( M_i^2\xi_+-2m^2),    \cr
{D}^{10}&=&  -   {\xi_-\over4}
\left[  28m^2-4M_i^2 \xi_+(3+\xi_+)      -
(\xi_+^2-\xi_-^2)Q^2 \right],   \cr
{D}^{00}&=&  + 2 {\xi_-\over3} 
[ -16m^2 + 2M_i^2 \xi_+(3+2\xi_+) 
\nonumber\\
&&\phantom{ + 2 {1\over3} 
\xi_-[ -16m^2}+  (\xi_+^2-\xi_-^2)Q^2].
\end{eqnarray}

A few remarks concerning the results of this section are in order:
\begin{itemize}
\item The equality (\ref{FeqFp}) follows from the equality
of (\ref{Nij}) and (\ref{barN_if}) at $Q^2=0$.
\item The elastic form factor (\ref{Fe_kkp}) is obtained from the
transition form factor (\ref{Fif_kkp}) setting $i=f$, and hence
$N^i_{tot}=N^f_{tot}=N_{tot}, M_i=M_f=M$ and $Q_c=0$. 

\item The normalization factors $N_{tot}^{i,f}$ which enter in Eqs.
(\ref{Fif_kkp}) are those of the corresponding   elastic form factors. 
The functions $g_n^0$ can be normalized arbitrarily  since any
multiplicative factor included in the definition of $g$ is cancelled
by a redefinition of $N_{tot}$. 
The function $g_2^1$ is, according to  Eq. (\ref{eq21c}), determined by
the choice for $g_2^0$.
We have used, for convenience, the normalization  $g_n^0(0)=1$.  

\item The transitions $n=2 \;\to \;n'=1$  are obtained by keeping only
the terms  $F^{00}$ and $F^{10}$ in (\ref{Fif_kkp}), as well as in the
corresponding expression $N_{tot}$.

\item The transitions $n=1 \;\to \;n'=2$  are similarly obtained by
keeping  the terms  $F^{00}$ and $F^{01}$. 

\item The transition $n=1 \;\to \;n'=1$ is given by the single term
$F^{00}$.
\end{itemize}


\end{document}